\documentclass[aps,twocolumn,pra,superscriptaddress,nofootinbib]{revtex4-2}
\pdfoutput=1

\usepackage{nicefrac}
\usepackage{amsmath}
\usepackage{amssymb}
\usepackage{amsthm}
\usepackage{mathrsfs}
\usepackage{bm, dsfont}
\usepackage[table]{xcolor}
\usepackage{graphicx}
\usepackage{natbib}
\usepackage[colorlinks=true,linkcolor=blue,citecolor=blue,urlcolor=blue]{hyperref}
\usepackage{cleveref}
\usepackage{verbatim, float}
\usepackage[normalem]{ulem}
\usepackage{physics}	
\usepackage[T1]{fontenc}
\usepackage[utf8]{inputenc}
\usepackage{multirow}
\usepackage{wrapfig}

\newcommand{\be}{\begin{equation}}
\newcommand{\ee}{\end{equation}}

\newcommand{\cE}{\mathcal{E}}

\newcommand{\ii}{\mathrm{i}}

\newcommand{\id}{\mathds{1}}
\newcommand{\cG}{\mathcal{G}}
\newcommand{\cI}{\mathcal{I}}

\newtheorem*{result}{Result}
\newcommand{\mean}[1]{\left \langle #1 \right\rangle}
\begin{document}

\title{Quantum sensing of a quantum field}

\author{Ricard Ravell Rodr\'iguez}
\affiliation{Institute for Cross-Disciplinary Physics and Complex Systems IFISC (UIB-CSIC),
Campus Universitat Illes Balears, E-07122 Palma de Mallorca, Spain}
\affiliation{ICFO-Institut de Ciències Fotòniques, The Barcelona Institute of
Science and Technology, 08860 Castelldefels (Barcelona), Spain}
\author{Mart\'i Perarnau-Llobet}
\affiliation{Física Teòrica: Informació i Fenòmens Quàntics, Department de Física,
Universitat Autònoma de Barcelona, 08193 Bellaterra (Barcelona), Spain}
\author{Pavel Sekatski}
\affiliation{Department of Applied Physics, University of Geneva, Switzerland}
\begin{abstract}
Estimating a classical parameter encoded in the Hamiltonian of a quantum probe is a fundamental and well-understood task in quantum metrology. A textbook example is the estimation of a classical field’s amplitude using a two-level probe, as described by the semi-classical Rabi model. In this work,  we explore the fully quantum analogue, where the amplitude of a coherent quantized field is estimated by letting it interact with a two-level atom. 
For both metrological scenarios, we focus on the quantum Fisher information (QFI) of the reduced state of the atomic probe.  
In the semi-classical Rabi model, the QFI is independent of the field amplitude and grows quadratically with the interaction time $\tau$. In contrast, when the atom interacts with a single coherent mode of the field, the QFI is bounded by 4, a constant dictated by the non-orthogonality of coherent states. We find that this bound can only be approached in the vacuum limit. In the limit of large amplitude $\alpha$, the QFI is found to attain its maximal value $1.47$ at $\tau =O(1)$ and $\tau =O(\alpha^2)$, and also shows periodic revivals at much later times. When the atom interacts with a sequence of coherent states, the QFI can increase with time but is bounded to scale linearly due to the production of entanglement between the atom and the radiation (back-action), except in the limit where the number of modes and their total energy diverge. Finally, in the continuous-field limit, where the atom interacts with a continuous source of weak coherent states, this back-action can be simply interpreted as spontaneous emission; we find that the optimal atomic QFI rate is finite, depends on the source intensity, and is upper bounded by the constant rate at which the QFI is emitted by the radiation source.

\end{abstract}
\maketitle

\section{introduction}
Quantum metrology is a field of research where the theoretically promised quantum advantage \cite{caves1981quantum,wineland1992spin,holland1993interferometric,giovannetti2004quantum,paris2009quantum} is practically explored on a wide range of platforms: gravitational wave detection \cite{aasi2013enhanced,demkowicz2013fundamental}, magnetometry \cite{wolfgramm2010squeezed,li2018quantum}, quantum plasmonic sensing \cite{dowran2018quantum}, or lossy phase shift estimation \cite{kacprowicz2010experimental}, to name a few. The textbook case study in quantum metrology is a two-level quantum probe interacting with a \emph{classical} field of unknown amplitude -- a scalar parameter in the Hamiltonian. 

We will specifically focus on the resonant (semi-classical) Rabi model~\cite{rabi1937space}, which describes the interaction of a two-level atom with a resonant harmonic electric field.  In the interaction picture with the rotating wave approximation, the Hamiltonian of the atom takes the well-known form
\be \label{eq: H class}
H =   E  g \, \ii \big(\ketbra{e}{g}-\ketbra{g}{e} \big),
\ee
where $E$ is the field amplitude, $g$ the coupling constant, and we chose the field quadrature such that the propagator is real. In this model, the evolution of the atom is simple -- its state undergoes the Rabi flops governed by the unitary propagator 
\begin{equation}
     V_{\tau} = e^{-\ii t H}=\exp( E \tau \big(\ketbra{e}{g}-\ketbra{g}{e}\big))
\end{equation}
where $\tau= t g $ is the interaction parameter that we simply refer to as the interaction time (here and below, we assume that the interaction time is perfectly known). After a certain time $\tau$, one can measure the probe in order to estimate the field amplitude $E$. The precision of the estimation is often quantified by the quantum Fisher information (QFI) \cite{braunstein1994statistical,paris2009quantum} of the evolved state $\rho_{\tau|E} = \ketbra{\Psi_{\tau|E}}$  with $\ket{\Psi_{\tau|E}}=V_{\tau}\ket{\Psi_0}$, see next section. In the case of a unitary evolution, the QFI takes a very simple form, and for $\ket{\Psi_0}=\ket{g}$ or $\ket{e}$, one finds 
\begin{equation}\label{eq: QFI Rabi}
    {\rm QFI}(\rho_{\tau|E}) = 4 \tau^2 = 4 g^2 t^2.
\end{equation}
Importantly, we see that $(i)$ it scales quadratically with time and $(ii)$ it is independent of the field amplitude $E$. Intuitively, this unlimited quadratic growth of the QFI with the interaction time is due to perfect coherence of the atomic evolution and absence of any back-action on the field. \\

The goal of this paper is to compare this simple analysis and its conclusions with a model where the atom interacts with a coherent quantized field of unknown amplitude. To do so, we first focus on the resonant Jaynes-Cummings model \cite{jaynes1963comparison,greentree2013fifty} where the atom interacts with a single mode of the electromagnetic field prepared in a coherent state.  In a second step, we consider a collision model~\cite{ciccarello2022quantum} built upon the Jaynes-Cummings interaction with a sequence of coherent modes, and its continuous limit. In both cases, we  derive  upper bounds on the QFI of the probe and processes approaching them.

Before moving forward, we note that the converse problem, namely the estimation of the atomic frequency through the emitted light, has been extensively studied in the literature, see e.g.~\cite{kiilerich2014estimation,kiilerich2016bayesian}. Other complementary previous works include the estimation of the coupling strength~\cite{genoni2012optimal} and the temperature~\cite{brunelli2012qubit} using the Jaynes-Cummings model.  

The article is structured as follows: in Section~\ref{sec: intro-QFI} we introduce the QFI as a benchmark for quantum sensing. Next, in Section~\ref{sec: JC model}, we study the interaction of the atom with a single mode of the field. Then in Section~\ref{sec: sequential}, we discuss the interactions with a sequence of field modes and the continuous limit. Finally, we present our conclusions.

\section{The quantum Fisher information, a benchmark for sensing}\label{sec: intro-QFI}

Consider the task of estimating a parameter $\theta$ by sampling a discrete random variable $X_\theta$, whose distribution $\bm p_\theta$ depends on the parameter. It can be analyzed by introducing an estimator $\hat \theta$ -- another random variable that is a function of $X_\theta$. It is natural to only consider estimators that are locally unbiased~\cite{berger2001statistical} (around $\theta_0$), i.e., for which the expected value equals the true value
$ \mathds{E}\left[\hat \theta(X_\theta)\right] = \theta$
for a small interval of parameter values around $\theta_0$. Furthermore, for an unbiased estimator, a natural quantifier of its precision is the associated mean-squared error 
$  {\rm MSE} (\hat \theta) =    \mathds{E}\left[\left(\hat \theta(X_\theta)-\theta\right)^2 \right].$
The renowned Cramér-Rao bound~\cite{rao1992information,cramer1999mathematical} relates the MSE of an unbounded estimator 
\be
{\rm MSE} (\hat \theta) \geq \frac{1}{{\rm FI}(X_\theta)}
\ee
to the Fisher Information (FI) of the random variable $X_\theta$. The latter is best understood as the susceptibility of $X_\theta$ with respect to several statistical distances, e.g. Kullback–Leibler divergence or Hellinger distance. In particular, the Fisher information is induced by the Euclidean norm on the normalized vector of probability ``amplitudes'' $\sqrt{\bm p_\theta}$ via
\begin{align}
{\rm FI}(X_\theta) &= 4  \left(\frac{\| \sqrt{\bm p_{\theta+\dd \theta}} -\sqrt{\bm p_{\theta}}\|}{\dd \theta}\right)^2 \\
& = 8 \, \left(\frac{1 - \sqrt{\bm p_{\theta+\dd \theta}} \sqrt{\bm p_{\theta}}}{\dd \theta^2}\right).
\end{align}

In the quantum estimation task we consider, the parameter is encoded into a state $\rho_\theta$ of the probe system. Thus, to define an estimator, one must first choose the measurement (POVM $\{ E_x \}$) to obtain the random variable $X_\theta$ distributed accordingly to the Born rule $p_{x|\theta}= \tr E_x \rho_\theta$. Nevertheless, one can show that optimizing over this measurement gives rise to the so-called Quantum Fisher Information (QFI) \cite{braunstein1994statistical,paris2009quantum} 
\begin{align}
    {\rm QFI}(\rho_\theta) &= \max_{\{ E_x \}} \, {\rm FI}(X_\theta)
    \\ \label{eq: qfi bures}
    & =8 \left(\frac{1- \|\sqrt{\rho_{\theta +\dd\theta}}\sqrt{\rho_\theta}\|_1}{\dd \theta^2}\right) 
\end{align}
with trace norm $\|A\|_1=\tr\sqrt{A^\dag A}$. Here, one recognizes the Uhlmann's fidelity $F(\rho,\sigma)=||\sqrt{\rho}\sqrt{\sigma} ||_1$ and the Bures distance $d_B(\rho,\sigma) = \sqrt{2(1-||\sqrt{\rho}\sqrt{\sigma} ||_1)}$, giving a geometric interpretation for the QFI. Furthermore, the QFI of a state can be computed~\cite{braunstein1994statistical,paris2009quantum} via 
\begin{align}\label{eq: QFI}
    {\rm QFI}(\rho_\theta) &= \Tr (L^2 \rho_\theta) \quad \text{for}\\
    \label{eq: SLD}
 \frac{L \rho_\theta + \rho_\theta L }{2}
    &= \frac{\dd \rho_\theta}{\dd \theta},
\end{align}
where $L$ is the so-called symmetric logarithmic derivative (SLD) defined by Eq.~\eqref{eq: SLD}. Note that some care has to be taken with this expression at parameter values where $\rho_\theta$ changes its rank and $L$ can become unbounded~\cite{ye2022quantum}.

By definition, the QFI bounds the mean-squared errors of all unbiased estimators obtained by measuring $\rho_\theta$, which is known as the quantum Crámer-Rao bound ${\rm MSE} (\hat \theta) \geq \frac{1}{{\rm QFI}(\rho_\theta)}$. For this reason, the QFI of the probe's state $\rho_\theta$ is commonly used as a figure of merit in quantum sensing, which is also the approach we take.

\subsection{Quantum Fisher Information of a two-level probe}

In this article, we focus on two-level probes, whose state can be expressed as  $\rho_\theta = \frac{1}{2}(\id + \bm r_\theta \cdot \tilde{\bm \sigma})$, where $\tilde {\bm \sigma} = (\sigma_x,\sigma_y,\sigma_z)$ is the vector of the three Pauli matrices. The QFI can easily be obtained (see Appendix \eqref{app:QFI-general} or~\cite{Zhong_2013} for an equivalent formula) as

\begin{equation}\label{eq: QFI-spin}
{\rm QFI}(\rho_\theta) = 
\frac{\|\dot{\bm r}_\theta\|^2 - \|\dot{\bm r}_\theta \cross \bm r_\theta\|^2}{1-\|\bm r_\theta\|^2 } =\|\dot{ \bm r}^\perp_\theta\|^2 + \frac{ \|\dot{ \bm r}^=_\theta \|^2}{1-\|\bm r_\theta\|^2}
\end{equation}
where $\| \bullet \|$ is the Euclidean norm, $\dot{\bm r}_\theta=\frac{\dd}{\dd \theta} \bm r_\theta$, and $\perp(=)$ denotes its orthogonal (parallel) component to $\bm r_\theta$. Note that the expressions on the right-hand side appear to diverge in the limit where $\rho_\theta$ is pure. We distinguish two cases. 

First, in the limit where $\|\bm r_\theta\|\to 1$ while $ \|\dot{ \bm r}^=_\theta \|$ does not vanish, the QFI is indeed unbounded. In fact, this is also the case for the FI of a random variable with distribution $\bm p_\theta = \binom{1-\theta}{\theta}$ at $\theta \to 0$. This case can be safely ignored for our study as nonphysical.

In contrast, when both $\|\bm r_\theta\|\to 1$ and $\|\dot{ \bm r}^=_\theta \| \to 0$, the QFI remains finite. It can be computed by either taking the limit~\cite{PhysRevA.95.052320} of the expression in Eq.~\eqref{eq: QFI-spin}, or by directly applying  Eq.~\eqref{eq: qfi bures} to obtain 
\begin{equation} {\rm QFI}(\rho_\theta)
= \| \ddot {\bm r}^=_{\theta}\| =\|\dot{\bm r}^\perp_\theta\|^2 
\end{equation}
with $\ddot {\bm r}_{\theta}= \frac{\dd^2}{\dd \theta^2}{\bm r}_\theta $. In this second case, the divergence is only apparent and is due to a coordinate singularity of pure states. Considering that in this case $\rho_\theta=\ket{\psi_\theta}\bra{\psi_\theta}$ one can also compute the QFI as
\begin{align}
    {\rm QFI}(\ket{\psi_\theta})&=4\left(\braket{\dot\psi_\theta}-\left|\bra{\dot \psi_\theta} \ket{\psi_\theta}\right|^2\right) \\
    &= 4\,\left\|\big(\id -\ketbra{\psi_\theta}\big)\ket{\dot \psi_\theta}\right\|^2
\end{align}


\section{Interaction of an atom with a single mode}\label{sec: JC model}

In this section, we study the problem of estimating the amplitude $\alpha$ of a single mode of the electromagnetic field, prepared in some coherent state $\ket{\alpha}=\sum_{n=0}^\infty \frac{\alpha^n}{\sqrt {n!}}e^{-\nicefrac{\alpha^2}{2}}\ket{n}$ with  $\alpha \in \mathbb{R}$, via a two-level atom probe coupled to the field with the resonant Jaynes-Cummings interaction. Contrary to the resonant Rabi model with a classical field (Eq.~\ref{eq: H class}),  where the QFI is independent of the field amplitude and grows quadratically with the interaction time (Eq.~\ref{eq: QFI Rabi}), we will see in this section that in the Jaynes-Cummings model the QFI has a complicated dependence on time and $\alpha$ and is always bounded by a constant.

\subsection{The QFI of coherent states themselves}

The first radical contrast with the estimation of a classical field in the Rabi model comes from the uncertainty principle. Indeed, the task of estimating $\alpha$ is intrinsically limited by the fact that coherent states are not orthogonal $|\braket{\alpha}{\beta}|= \exp(-\frac{|\alpha-\beta|^2}{2})$. In particular, by Eq.~\eqref{eq: qfi bures} the QFI of the coherent states (with respect to the amplitude parameter $\alpha$) is given by  
\be\label{eq: QFI coh}
{\rm QFI}\Big(\ketbra{\alpha}\Big)= 8 \frac{1-|\braket{\alpha}{\alpha+\dd\alpha}|}{\dd \alpha^2} =4.
\ee
Now, let $\rho_{\tau|\alpha}$ be the state of the two-level atom after an interaction with the coherent field mode $\ket{\alpha}$. Since QFI is monotonic (can not increase by post-processing), Eq.~\eqref{eq: QFI coh} immediately implies
\be\label{eq: 4 UB}
{\rm QFI}\Big(\rho_{\tau|\alpha} \Big) \leq 4,
\ee
showing that the QFI of the probe is bounded for all amplitudes $\alpha$ of the field, and all interaction times $\tau$.\\

The interaction between the field and the atom is mapping the information encoded in the infinite-dimensional Hilbert space of the field mode into the qubit carried by the atom. In particular, the bound~\eqref{eq: 4 UB} can always be saturated by swapping the two-dimensional subspace spanned by $\ket{\alpha}$ and $\ket{\dot \alpha}=\sum_{n=0}^\infty \frac{\alpha^n (n-\alpha^2)}{\alpha\sqrt {n!}}e^{-\nicefrac{\alpha^2}{2}}\ket{n}$, with $\braket{\dot \alpha}=1$ and $\braket{\alpha}{\dot \alpha}=0$, into the probe; e.g. with the SWAP unitary 
\be\label{eq: V opt}
V = \id +
\ketbra{g,\dot \alpha}{e,\alpha} + \ketbra{e, \alpha}{g,\dot \alpha}- \ketbra{g,\dot \alpha} -\ketbra{e, \alpha}.
\ee
As we show below, in the vacuum limit $\alpha\ll 1$ the Jaynes-Cummings interaction attains the theoretical upper-bound by realizing $V$ in Eq.~\eqref{eq: V opt}. Nevertheless, this interaction is fine-tuned to work around a specific value of $\alpha$. In general, there exists no {\it fixed} atom-field interaction saturating ${\rm QFI(\rho_{\tau|\alpha})}=4$ {\it for all} $\alpha$~\cite{inprep}.  Quite remarkably, as we also show, for $\tau=1$ the Jaynes-Cummings interaction gives a decent ${\rm QFI}(\rho_{\tau|\alpha})= \nicefrac{4}{e}\approx 1.47$ for all $\alpha\gg 1$ (see Fig.~\ref{fig:tau1}). It is an interesting open question to identify the interaction $V$ that maximizes the worst-case value $\min_\alpha {\rm QFI}(\rho_{\tau|\alpha})$. We conjecture that the optimal value is $\nicefrac{4}{e}$, or very close to it, as suggested by another natural interaction model discussed in Sec.~\ref{sec: covariant interaction}.

\subsection{The resonant Jaynes-Cummings model}

Next, we consider the resonant Jaynes-Cummings model where the two-level atom interacts with a single mode of the electromagnetic field, with associated creation and annihilation operators $a^\dag, a$ \cite{jaynes1963comparison,gerry2023introductory}. In the interaction picture, the Jaynes-Cummings Hamiltonian is given by
\be
H = \ii g \left( \sigma_+ a - \sigma_- a^\dag \right).
\ee
Assuming that the field and the atom are initially pure states $\ket{\mathfrak{F}}$ and $\ket{\Psi_0}$, their joint state after interaction time $\tau= tg$ becomes 
\be\label{eq: prop JC}
U_\tau \ket{\Psi_0} \ket{\mathfrak{F}} \quad \text{for } \quad U_\tau = \exp(\tau \left( \sigma_+ a - \sigma_- a^\dag \right)).
\ee

It is a standard exercise in quantum optics to see that in the computational basis of the atom $\{\ket{g},\ket{e}\}$ the propagator in Eq.~\eqref{eq: prop JC} can be expressed as (see e.g. \cite{sekatski2010cloning} for an explicit calculation)
\begin{align}
 U_\tau &= 
 \left(\begin{array}{cc}
 \cos(\tau \sqrt{\hat n}) & -  \frac{\sin(\tau \sqrt{\hat n })}{\sqrt{\hat n }} a^\dag \\
  \frac{\sin(\tau \sqrt{\hat n+1})}{\sqrt{\hat n+1}} a & \cos(\tau \sqrt{\hat n+1}) 
  \end{array}\right),
\end{align}
where $\hat n=a^\dag a$ is the photon number operator. The reduced dynamics of the atom is then given by the completely positive trace preserving (CPTP) map $\cE_{\tau|\mathfrak{F}}[\bullet]=\tr_{\rm field} U_\tau \left(\bullet\otimes \ketbra{\mathfrak{F}} \right) U_\tau^\dag$. Straightforward algebra allows one to write
\be \label{eq: channel in Gramm}
\cE_{\tau|\mathfrak{F}}[\bullet] =\sum_{i,j=0}^3 G^\mathfrak{F}_{ij} \, L_i \bullet L_j^\dag, 
\ee
where $\{L_0,L_1,L_2,L_3\} =\{ \ketbra{g},\ketbra{e},\sigma_-,\sigma_+\}$ and $G^{\mathfrak{F}}_{ij}=\braket{\mathfrak{F}_j}{\mathfrak{F}_i}$, termed the channel-matrix, is the Gram matrix of the unnormalized states
 \begin{equation}\label{eq: state gens}
 \begin{split}
\ket{\mathfrak{F}_0} &=\cos(\tau \sqrt{\hat n })\ket{\mathfrak{F}},\quad \qquad  \ket{\mathfrak{F}_2} =- \frac{\sin(\tau \sqrt{\hat n })}{\sqrt{\hat n}} a^\dag\ket{\mathfrak{F}} ,
\\
\ket{\mathfrak{F}_1} &=\cos(\tau \sqrt{\hat n+1 })\ket{\mathfrak{F}},\quad
\ket{\mathfrak{F}_3} = \frac{\sin(\tau \sqrt{\hat n+1})}{\sqrt{\hat n+1}} a \ket{\mathfrak{F}}.
\end{split}\nonumber
 \end{equation} 
By construction the Gram matrix is Hermitian $G^{\mathfrak{F}}_{ji} = G^{\mathfrak{F}*}_{ij}$. In addition, the fact that the map is trace preserving implies the equality constraints $G_{00}^{\mathfrak{F}}+G_{33}^{\mathfrak{F}}=G_{11}^{\mathfrak{F}}+G_{22}^{\mathfrak{F}}=1$ and $G_{02}^{\mathfrak{F}}+G_{31}^{\mathfrak{F}}=0$. The Kraus operators for the channel $\cE_{\tau|\mathfrak{F}}$ can be obtained by diagonalizing the matrix $G^{\mathfrak{F}}$, however a more intuitive representation of the dynamics of the two-level atom is given by the induced affine transformation of the Bloch vector $\cE_{\tau|\mathfrak{F}}[\bm \sigma]$ with $\bm \sigma = (\id,\sigma_x,\sigma_y,\sigma_z)$.

\subsection{The atomic dynamics and QFI}

\begin{figure*}
\newcommand{\HG}{0.169}
    \centering
     \includegraphics[height=\HG\linewidth]{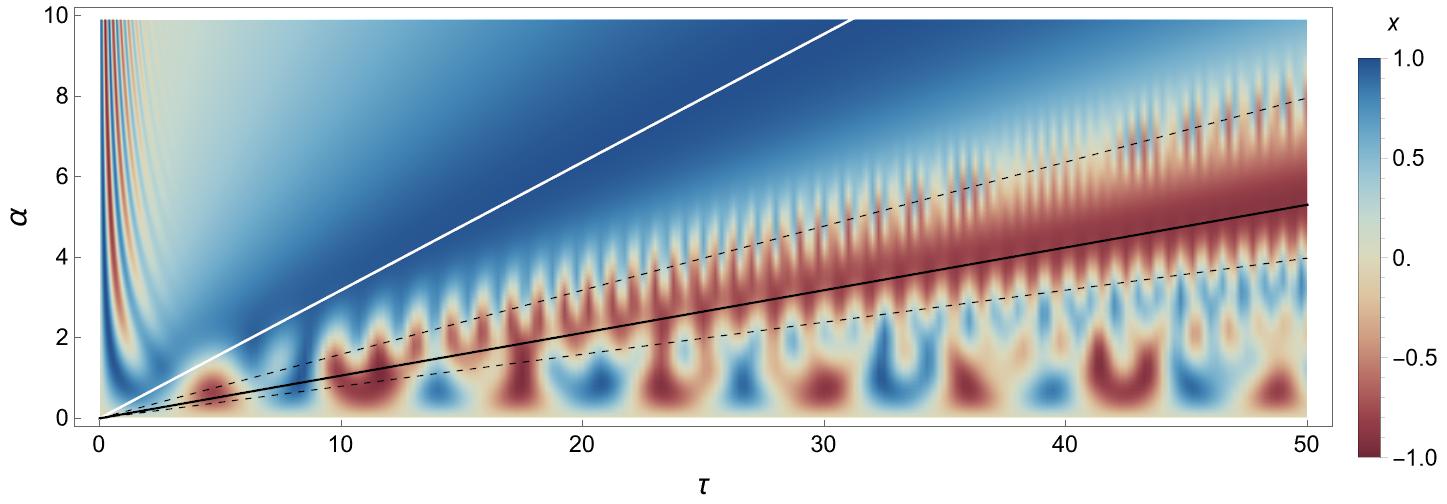}
       \includegraphics[height=\HG\linewidth]{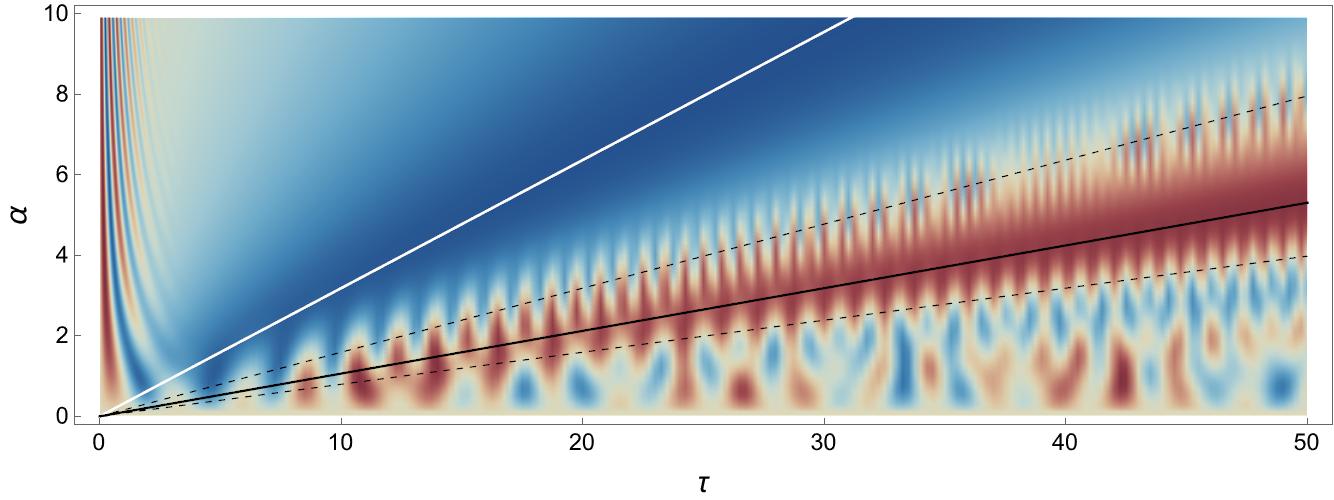}
    \includegraphics[height=\HG \linewidth]{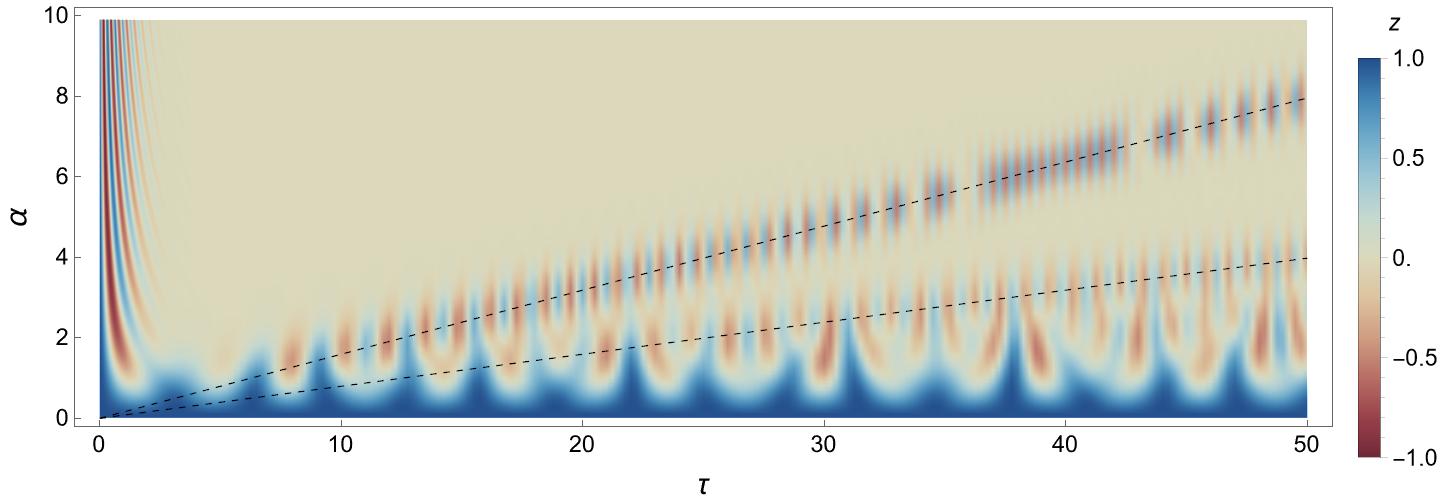}
    \includegraphics[height=\HG\linewidth]{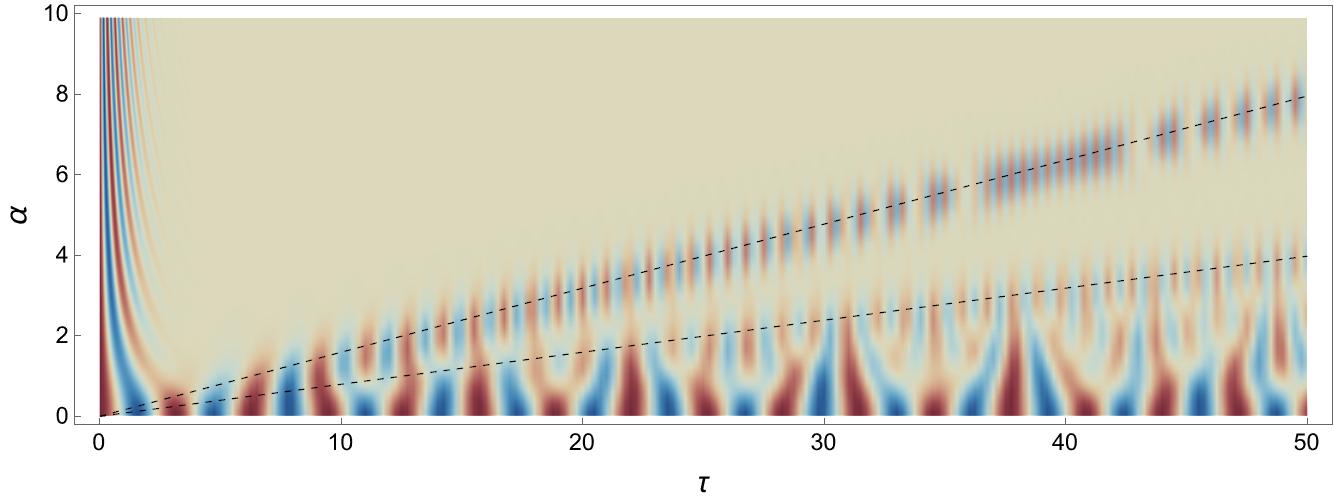}   \hspace{1 mm}
    \caption{From top to bottom: the coherence $x_{\tau|\alpha}^{(g/e)}$ and the population $z_{\tau|\alpha}^{(g/e)}$ of the atomic state as function of time $\tau$ (x-axis) and coherent state amplitude $\alpha$ (y-axis). The left column corresponds to the ground state, and the right to the excited. The guidelines represent the different ``revival'' times derived analytically in the asymptotic limit (see Appendix~\ref{app: large alpha limit}).}
    \label{fig: coh-pop}
\end{figure*}

 We now let the field be in a coherent state $\ket{\mathfrak{F}}=\ket{\alpha}$, and investigate how well its amplitude $\alpha$ can be estimated from the atomic probe. To do so, we focus on the QFI of the reduced state of the atom
 \begin{align}\label{eq: QFI pure}
 \rho_{\tau|\alpha} = \cE_{\tau|\alpha}\left[\rho_0\right].
 \end{align}
In this case, using the form of the coherent state, the states $\ket{\mathfrak{F}_i}=\ket{\alpha_i}$ in Eq.~\eqref{eq: state gens} can be expressed as $\ket{\alpha_i}=f_i(\hat n) \ket{\alpha}$ (see App.~\ref{app: cutoff comp}), with 
\begin{equation}\label{eq: function f}
 \begin{split}
f_0(\hat n) &=\cos(\tau \sqrt{\hat n }), \quad \qquad  f_2(\hat n) = - \frac{\sqrt{\hat n}}{\alpha}\sin(\tau \sqrt{\hat n })  ,
\\
f_1(\hat n) &=\cos(\tau \sqrt{\hat n+1 }),\quad
f_3(\hat n)= \frac{\alpha}{\sqrt{\hat n+1}} \sin(\tau \sqrt{\hat n+1}).
\end{split}
 \end{equation} 
This allows us to compute the elements of the Gram matrix 
$G^\alpha_{ij} 
= \bra{\alpha} f_j(\hat n) f_i(\hat n) \ket{\alpha} = \sum_{n=0}^\infty {\rm P}(n|\alpha) f_i(n) f_j(n)$
as expected values of simple functions of a Poisson random variable 
${\rm P}(n|\alpha)=|\braket{n}{\alpha}|^2=\frac{\alpha^{2n}}{n!}e^{-\alpha^2}$. In turn, using $\dot {\rm P}(n|\alpha) = 2\frac{n-\alpha^2}{\alpha} {\rm P}(n|\alpha)$,  we can also express the derivatives $\dot G_{ij}^\alpha=\frac{\dd }{\dd\alpha } G_{ij}^\alpha$ in the same form.

For concreteness, let us now focus on the case where the atomic probe is initialized in the ground or excited state. After time $\tau$ its state remains real 
\be\label{eq: g and e}
\rho_{\tau|\alpha}^{(g/e)} =\frac{1}{2}\left(\id + x_{\tau|\alpha}^{(g/e)}\sigma_x + z_{\tau|\alpha}^{(g/e)}\sigma_z \right).
\ee
The QFI of the final state can be computed with the help of the Eq.~\eqref{eq: QFI-spin},  which simplifies to
\begin{equation}\label{eq: QFI xz}
{\rm QFI}[\rho_{\tau|\alpha}]
=\frac{\dot x_{\tau|\alpha}^2+\dot z_{\tau|\alpha}^2 - (\dot x_{\tau|\alpha} z_{\tau|\alpha} - \dot z_{\tau|\alpha} x_{\tau|\alpha})^2}{1-x_{\tau|\alpha}^2-z_{\tau|\alpha}^2}.
\end{equation}
The quantities $x_\alpha$ with $z_\alpha$ and their derivatives $\dot x_\alpha$ with $\dot z_\alpha$  are linear combinations of $G_{ij}^\alpha$ and $\dot G_{ij}^\alpha$ and can thus be expressed as expected values of certain functions $\mathds{E}[g(\hat n)]$ of the photon number $\hat n$ -- a Poisson random variable.  In particular, for the ground state, we find
\begin{align} \label{eq: main xg}
    x_{\tau|\alpha}^{(g)} &= \,\mathds{E}\left[{\frac{ 2\alpha}{\sqrt{\hat n +1}} \cos(\tau\sqrt{\hat n}) \sin(\tau\sqrt{\hat n+1}) \, } \right], \\ \nonumber
    z_{\tau|\alpha}^{(g)} &= \,\mathds{E}\left[2 \cos^2(\tau\sqrt{\hat n})\right]-1, \\ \nonumber
\dot x_{\tau|\alpha}^{(g)} &=
 \, \, \mathds{E}\left[{  \frac{ 4\hat n-4\alpha^2+2 }{\sqrt{\hat n+1}}}\,\cos(\tau\sqrt{\hat n}) \sin(\tau\sqrt{\hat n+1}) \right],\\ \nonumber
\dot z_{\tau|\alpha}^{(g)} &= \,\mathds{E}\left[{ \frac{4\hat n-4\alpha^2}{\alpha} \cos^2(\tau\sqrt{\hat n})}\right],
\end{align} 
which are derived in App.~\ref{app: cutoff comp} together with similar expressions for the excited initial state.

Although in general, we cannot compute these expected values analytically (the limits of small and big $\alpha$ are discussed in the following sections), these are well-behaved sums that can be well approximated numerically, e.g., using a cutoff on the photon number. The same holds for the QFI, provided that the state is not pure, so that the denominator in
Eq.~\eqref{eq: QFI xz} remains bounded from zero. This can be verified by computing the purity of the state
\be
\mathcal{P}[\rho_{\tau|\alpha}]:= 2 \tr \rho_{\tau|\alpha}^2-1 = x_{\tau|\alpha}^2 + z_{\tau|\alpha}^2,
\ee
where the last equality assumes a real qubit state. Since the global dynamics is unitary, and $1-\mathcal{P}$ is a Schur-convex function of the eigenvalues of $\rho_{\tau|\alpha}$, it is also a measure of atom-field entanglement.

\begin{figure*}[t]
\newcommand{\HG}{0.169}
    \centering
     \includegraphics[height=\HG\linewidth]{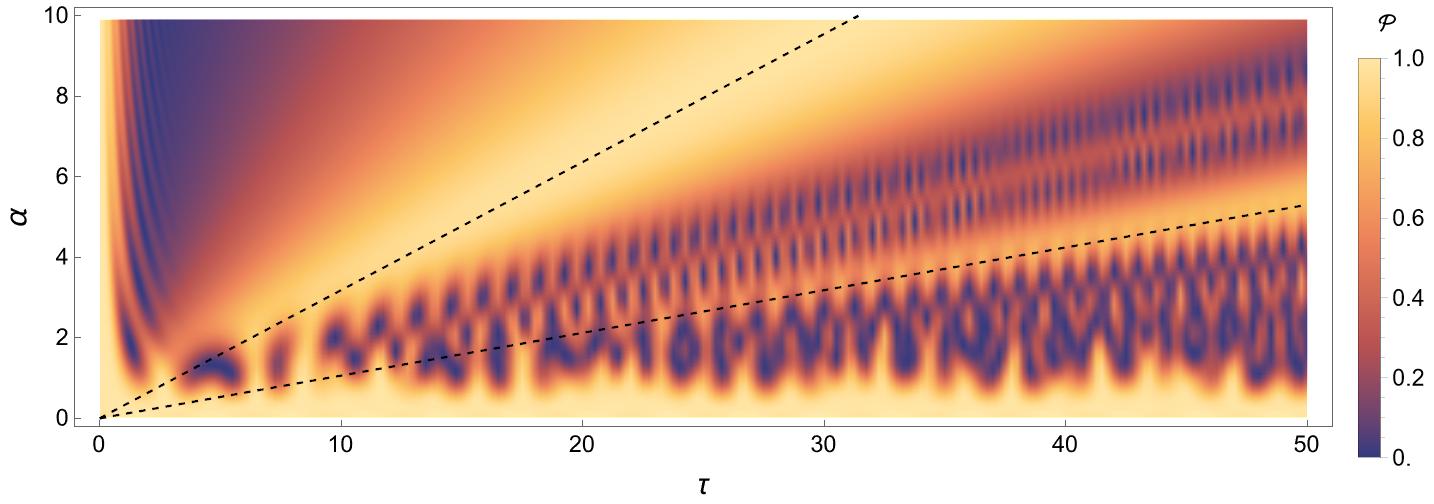}
       \includegraphics[height=\HG\linewidth]{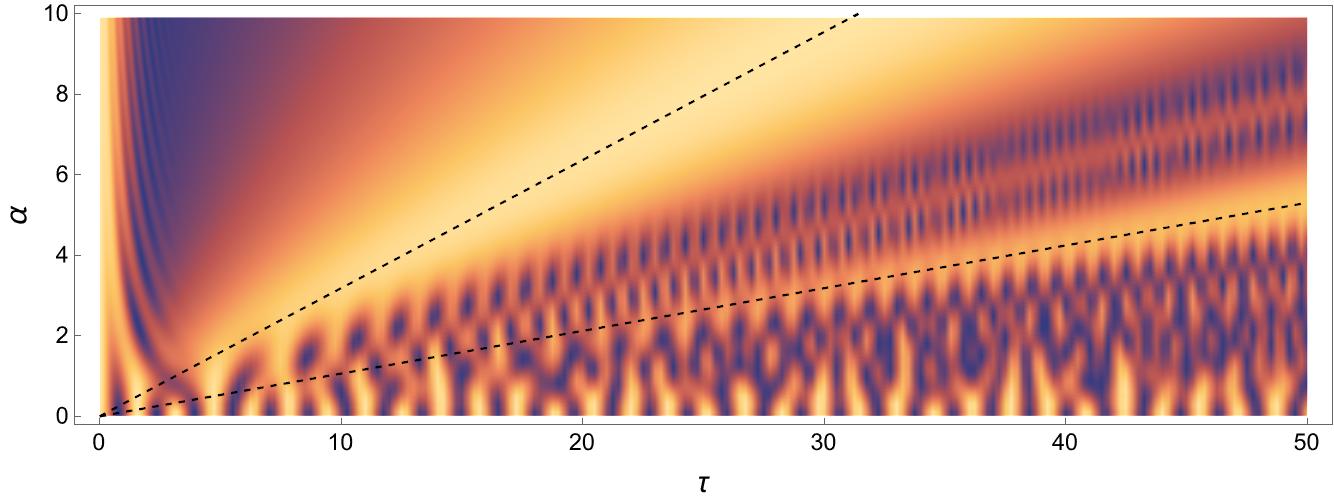}
   \includegraphics[height=\HG \linewidth]{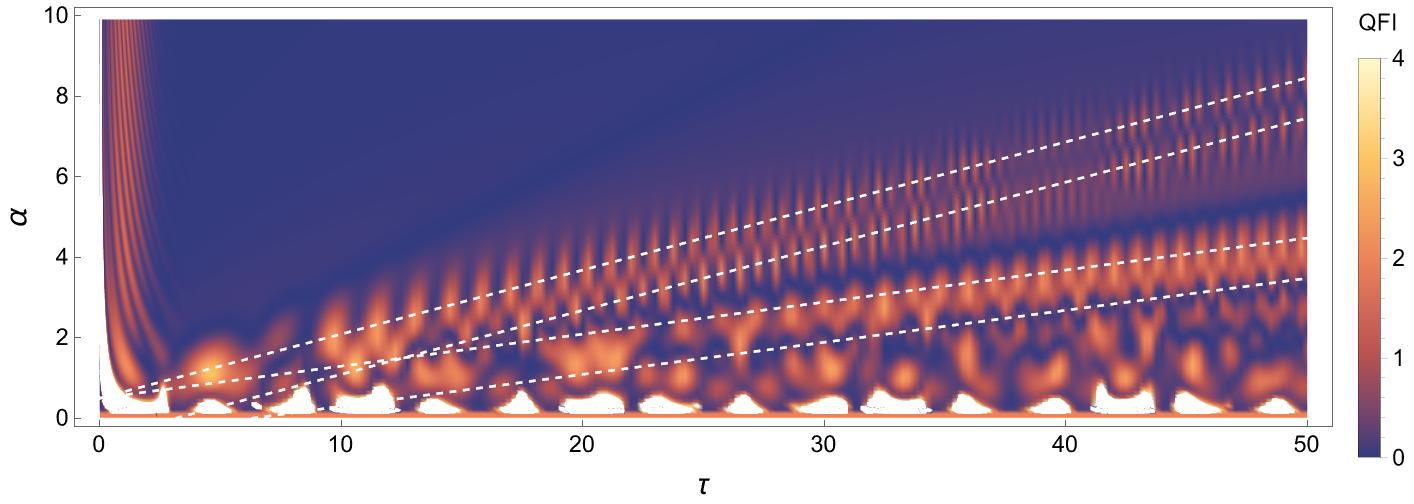}
   \hspace{1 mm}
   \includegraphics[height=\HG\linewidth]{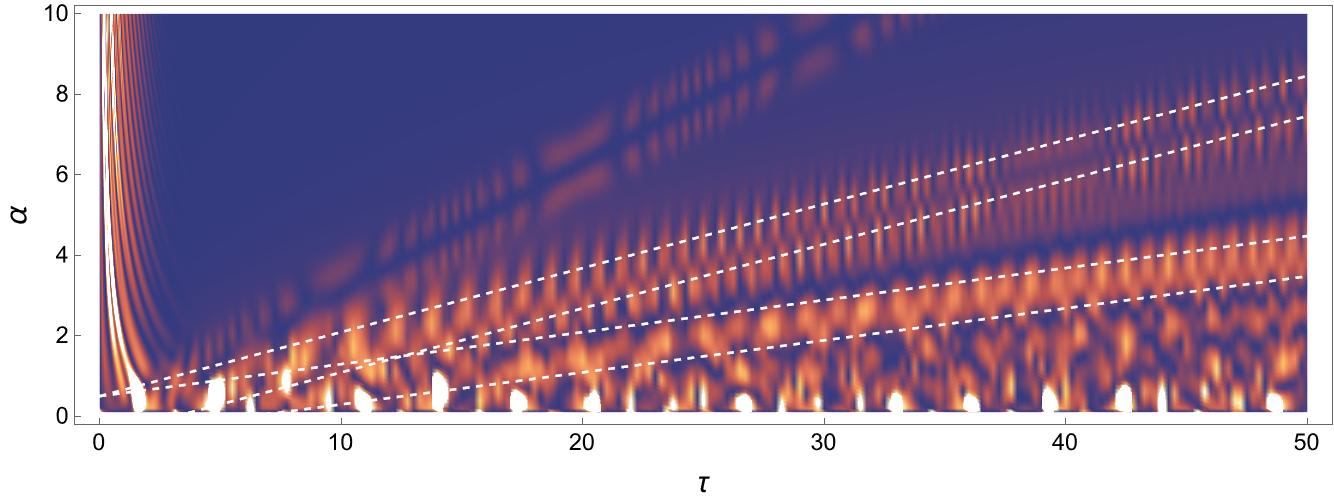}
    \caption{From top to bottom: the purity $\mathcal{P}[\rho_{\tau|\alpha}^{(g/e)}]$ and the quantum Fisher information ${\rm QFI}[\rho_{\tau|\alpha}^{(g/e)}]$ of the atomic state as function of time $\tau$ (x-axis) and coherent state amplitude $\alpha$ (y-axis). The left column corresponds to the ground state, and the right to the excited. The guidelines represent the different ``revival'' times derived analytically in the asymptotic limit (see Appendix~\ref{app: large alpha limit}). $1-\mathcal{P}[\rho_{\tau|\alpha}^{(g/e)}]$ measures the atom-field entanglement. The white patches that appear for small values of $\alpha$ in the bottom figures are due to the state being pure close to $\alpha = 0$; it is a numerical artifact.}
    \label{fig: qfiground}
\end{figure*}

The results for the atomic coherence $ x_\alpha^{(g/e)}$ and the population $z_\alpha^{(g/e)}$ for $\alpha \in[0,10]$ and $\tau\in[0,50]$ are reported in Fig.~\ref{fig: coh-pop},  while similar results for the purity and QFI of the final atomic state are reported in Fig.~\ref{fig: qfiground}. We now briefly discuss these numerical results before diving into a detailed analysis of different limits in the following sections. 

At very low $\alpha$ (Sec.~\ref{sec:vacucm limit}) the ground state remains unchanged while the excited atom is subject to vacuum Rabi oscillations. We see that in both cases close to the vacuum field $\alpha\approx 0$, the QFI has a quasi-periodic dependence on time, and approaches the maximal possible value of $4$.

At relatively short times $\tau\approx O(1)$ (Sec.~\ref{sec: short t}), the atomic state has a periodic dependence on $\alpha$. Here, the QFI is also non-zero. However, we find that in the asymptotic limit $\alpha\gg1$ it never exceeds $1.47$.  

Later, at $\tau\gg 1$ (Sec.~\ref{sec: main limit}), the atomic population and the QFI quickly vanish, while the atomic coherence shows slow oscillations. Remarkably, at specific times (full white and black lines in Fig.~\ref{fig: coh-pop}), the atom becomes pure, with $x_{\tau|\alpha}^2\approx 1$, 
and independent of the initial state~\cite{gea1990collapse,Zhang2015quantum}, 
When time further increases (dashed lines in Fig.~\ref{fig: coh-pop}), the atomic population goes through sharp sudden revivals~\cite{eberly1980periodic,rempe1987observation,phoenix1988fluctuations,gea1990collapse,fleischhauer1993revivals}. We see that the QFI follows the same trend and can increase up to 0.54 (around the first revival).

Interestingly, for even longer times $\tau\gg \alpha$ (Sec.~\ref{sec: main limit} and \ref{sec: long t}) we find that the QFI can reach even higher values while the atomic state is not fully mixed until $\tau \gg \alpha^3$.

The majority of the remainder of the section is devoted to studying the two aforementioned limits: the vacuum limit where $\alpha$ is small, and the large $\alpha$ limit. Finally, we briefly investing using encodings beyond the JC model.


\subsection{The vacuum limit $\alpha \ll 1$}
\label{sec:vacucm limit}
We start by examining the regime of a very weak field $\alpha \ll 1$. Recall that at $\alpha = 0$ one recovers the famous vacuum Rabi oscillations, with the reduced dynamics of the atom given by the spontaneous-decay channel
\begin{align}
\cE_{\tau|0}[\bullet] &= K_0 \bullet K_0^\dag + K_1\bullet K_1^\dag \quad \text{with} \\
K_0 &=\left(\begin{array}{cc} 
0 & \sin(\tau)\\
0& 0
\end{array}\right), \quad
K_1 = \left(\begin{array}{cc} 
1 & 0\\
0& \cos(\tau)
\end{array}\right).
\end{align}
To compute the QFI of the atomic state, we need the expression of the channel $\cE_{\tau|\alpha}$ and its derivative up to the second order in $\alpha$. It can be computed using $\ket{\alpha}\propto \ket{0}+\alpha \ket{1} +\nicefrac{\alpha^2}{2}\ket{2}$ (see Appendix~\ref{app: vacuum limit}). In particular, for the atom initially prepared in the ground or excited state, we find
\begin{align}
\rho_{\tau|\alpha}^{(g)} &=
\left(\begin{array}{cc} 
1-{\rm s}^2_\tau \alpha^2 & \alpha\,  {\rm s}_\tau\\ 
\alpha\,  {\rm s}_\tau & \alpha^2 {\rm s}^2_\tau
\end{array}\right),\\
\rho_{\tau|\alpha}^{(e)} &=
\left(
\begin{array}{cc}
{\rm s}^2_\tau + \alpha ^2 \left({\rm s}_{\sqrt 2 \tau}^2-{\rm s}^2_\tau\right) & 
 -\alpha \,  {\rm s}_\tau {\rm c}_{\sqrt{2}\tau} \\
 -\alpha  \, {\rm s}_\tau {\rm c}_{\sqrt{2} \tau} & {\rm c}_\tau^2 + \alpha ^2 \left({\rm c}_{\sqrt 2\tau}^2- \rm{c}_\tau^2 \right) \\
\end{array}
\right),
\end{align}
where ${\rm s}_{\gamma}= \sin(\gamma)$, ${\rm c}_\gamma = \cos(\gamma)$. Using Eqs.~\eqref{eq: main xg} and \eqref{eq: QFI xz}, it is also direct to compute the QFI
\begin{align}\label{eq: QFI vaccum g}
{\rm QFI}[\rho_{\tau|\alpha}^{(g)}] &= 4 \sin^2(\tau) +O(\alpha^2), \\
\label{eq: QFI vaccum e}
{\rm QFI}[\rho_{\tau|\alpha}^{(e)}] &= 4 \sin ^2(\tau ) \cos ^2\left(\sqrt{2} \tau \right) +O(\alpha^2).
\end{align}
 Remarkably, in both cases we see that the QFI periodically approaches the maximum value of $4$ at some well-chosen times $\tau$, reminiscent of Rabi oscillations. In fact, this is the case for all real initial states $\cos(\theta)\ket{g}+\sin(\theta)\ket{e}$ of the atom.

\subsection{The large  $\alpha \gg 1$ limit }
\label{sec: main limit}

A very crude ``classical field approximation'' that one can sometimes encounter consists of saying that for large enough $\alpha$ one has $a^\dag \ket{\alpha} \approx \alpha \ket{\alpha}$ in addition to $a\ket{\alpha}=\alpha\ket{\alpha}$. Replacing the creation $a^\dag$ and annihilation $a$ operators with the scalar $\alpha$ in the Hamiltonian gives rise to the approximation $H \approx g \, \alpha \, \ii \big(\ketbra{e}{g}-\ketbra{g}{e} \big)$ reproducing the semi-classical Rabi model in Eq.~\eqref{eq: H class}. As we will see, this approximation is only accurate in the limit $\tau\ll 1$. Instead, we now derive the exact reduced evolution of the atom.

The Jaynes-Cummings interaction with a coherent state in the large $\alpha$ limit has been extensively studied in the literature, see e.g. Refs.~\cite{cummings1965stimulated,eberly1980periodic,rempe1987observation,phoenix1988fluctuations,gea1990collapse,PhysRevA.44.5913,buvzek1992schrodinger,shore1993jaynes,fleischhauer1993revivals,azuma2010application,berman2014collapse,Zhang2015quantum,pavlik2023inside}. Our goal here is modest, as we are only interested in the reduced dynamics of the atom. Despite the large body of literature, we could not find a reference describing the full reduced dynamics of the atom at all times. Hence, we begin this section by deriving such an expression. We will then interpret the obtained result and discuss the atomic QFI in different time regimes.

\begin{result} 
    In the large $\alpha\gg 1$ limit, the atomic dynamics induced by the resonant Jaynes-Cummings interaction with a coherent field is at all times  given by 
    \be\label{eq: channel general}
    \cE_{\tau|\alpha} = \mathcal{T}_{\nicefrac{\id}{2}} + \sum_{\nu=0}^\infty \left(\mathcal{R}_{Q_\nu,\Phi_\nu}^\nu + \mathcal{D}_{P_\nu,\Omega_\nu}^\nu\right),
    \ee
    where $\mathcal{T}_{\nicefrac{\id}{2}}[\bm \sigma]=(\id,0,0,0)$ is the totally depolarizing channel, the linear maps are given by
    \begin{align}
     \mathcal{R}_{q,\varphi}^\nu[\bm \sigma]& :=\left(
    \begin{array}{c}
    0\\
    q \big((-1)^\nu \cos(\varphi) \sigma_x - \sin(\varphi)\sigma_z\big)
    \\
   0
    \\
    q \big((-1)^\nu \sin(\varphi) \sigma_x + \cos(\varphi)\sigma_z\big)
    \end{array}
    \right),\label{eq: R channel} \\
    \mathcal{D}^{\nu}_{p,\omega}[\bm \sigma]&:=
     \left(
    \begin{array}{c}
    p (-1)^\nu \sin(\omega) \sigma_x\\
     0
    \\
    p (-1)^\nu \cos(\omega) \sigma_y 
    \\
     0
    \end{array}
    \right), \label{slow phase}
\end{align}
and the functions $Q_\nu(\tau|\alpha), \Phi_\nu(\tau|\alpha), P_\nu(\tau|\alpha)$ and $\Omega_{\nu}(\tau|\alpha)$ are 
given in Eqs.~(\ref{eq: R0}-\ref{eq: Omega0}) for $\nu=0$ and Eqs.~(\ref{eq: R_nu}-\ref{eq: omega_nu}) for $\nu\geq 1$. 
\end{result}

\begin{proof}
To understand the atomic dynamics  in  the large amplitude limit, it is insightful to observe that the entries of the channel-matrix
\be\label{eq: GGG}
G^\alpha_{ij} = \sum_{n=0}^\infty {\rm P}(n|\alpha) f_i(n) f_j(n),
\ee
with $f_i(n)$ in Eq.~\eqref{eq: function f}, result from the interference of terms oscillating on two separate time scales -- with the rapid  $\widetilde S(n)\in \{2 \tau \sqrt{n}, \tau (\sqrt{n}+\sqrt{n+1}),2 \tau \sqrt{n+1}\}$ and the slow $\bar S(n)=\tau (\sqrt{n}-\sqrt{n+1})$ phases. Using $\frac{\sqrt {\hat n}}{\alpha},\frac{\alpha}{\sqrt{\hat n+1}} =1 +O(\nicefrac{1}{\alpha})$ we can thus decompose the channel-matrix as 
\be
G = \nicefrac{\id}{2} + \widetilde \cG +\bar \cG+O(\nicefrac{1}{\alpha}) 
\ee
where each term $\cG$ is a linear combination of terms $\mathds{E}[e^{\ii S(n)}] :=\sum_n P(n|\alpha) e^{\ii S(n)}$ (and their Hermitian conjugates) for the rapid and slow phases, respectively. By linearity, the same holds for the channel, which in the large $\alpha$ limit reads
\be
\cE_{\tau|\alpha} = \mathcal{T}_{\nicefrac{\id}{2}}+\widetilde \cE_{\tau|\alpha} + \bar\cE_{\tau|\alpha}.
\ee
Here, $\mathcal{T}_{\nicefrac{\id}{2}}[\bm \sigma]=(\id,0,0,0)$ is the totally depolarizing channel coming from $G=\nicefrac{\id}{2}$, it maps any input state to the maximally mixed one $\mathcal{T}_{\nicefrac{\id}{2}}[\rho_0]=\nicefrac{\id}{2}$. The linear maps $\widetilde \cE_{\tau|\alpha} $ and $\bar\cE_{\tau|\alpha}$, coming from rapid and slow phase terms, describe very different dynamics and manifest at different time scales, as we now show. All the technical details underlying the following discussion can be found in Appendix~\ref{app: large alpha limit}. \\

For relatively short times, the phases can be linearized. Indeed in the large $\alpha$ limit the Poisson distribution concentrates in a width of $O(\alpha)$, so the rescaled random variable  satisfies $\hat \delta = \frac{\hat n -\alpha^2}{\alpha}=O(1)$, implying 
\begin{align}
\widetilde S(n)&= 2 \alpha \tau + \tau\, \delta +O(\nicefrac{\tau}{\alpha}),\\
\bar S (n) &= -\frac{\tau}{2\alpha} +\frac{\tau}{4 \alpha^2} \delta +O(\nicefrac{\tau}{\alpha^3}).
\end{align}
Using the moment generating function of the Poissonian $\mathds{E}[e^{\ii \mu n}]  = \exp(-\alpha ^2+\alpha ^2 e^{i \mu })$ we directly obtain the expressions of the maps  
\begin{align} \label{eq: R0}  
\widetilde \cE_{\tau|\alpha} &= \mathcal{R}_{Q_0,\Phi_0}^0, &\quad Q_{0}(\tau|\alpha)& := e^{-\nicefrac{\tau^2}{2}}\\
&&
\Phi_{0}(\tau|\alpha) &:= 2\alpha\, \tau
\end{align}
valid in the limit $\nicefrac{\tau}{\alpha}\ll1$, and
 \begin{align}
    \bar \cE_{\tau|\alpha} &= \mathcal{D}^{0}_{P_0,\Omega_0} &\quad 
   P_0(\tau|\alpha)&:= e^{-\nicefrac{\tau^2}{32\alpha^4}} 
   \\
   && \Omega_0(\tau|\alpha)& := \nicefrac{\tau}{2\alpha}
    \label{eq: Omega0}
 \end{align}
valid for $\nicefrac{\tau}{\alpha^3}\ll1$, respectively.\\

In contrast, for longer times $\tau \geq O(\alpha)$, resp. $\tau \geq O(\alpha^3)$, when these approximations break down, the phases $\widetilde S(n)$, resp. $\bar S(n)$, oscillate rapidly as compared to the envelope $P(n|\alpha)$. Hence, following~\cite{fleischhauer1993revivals}, we can compute  $\mathds{E}[e^{\ii S(n)}]$ in two steps. First, using the Poisson summation formula, we replace the discrete sum with a series of integrals 
\be
\mathds{E}[e^{\ii S(n)}] = \sum_{\nu=-\infty}^\infty \int_0^\infty\!\! \dd n \,  {\rm P}(n|\alpha) e^{\ii S(n)}e^{-\ii 2\pi \nu n},
\ee
where we neglected the boundary term $\frac{1}{2}{\rm P}(0|\alpha) e^{\ii S(0)}$.
Second, we use the stationary phase approximation to evaluate the integrals. As a result, we find that the phases only interfere constructively around well-separated times:  respectively $\tau \approx 2\pi \alpha \nu$ and $\tau \approx 8 \pi \nu \alpha^3$ with positive integer $\nu\geq 1$. This leads to sharp revivals of the short-time dynamics, with
\begin{align}
&\widetilde \cE_{\tau|\alpha} = \mathcal{R}_{Q_\nu,\Phi_\nu}^\nu, &\quad
   Q_\nu(\tau|\alpha) &:=  \frac{e^{-\left(\frac{\tau-2\pi\alpha\nu}{\sqrt 2\pi \nu}\right)^2}}{\sqrt{\pi \nu}}\label{eq: R_nu} \\
   && \Phi_\nu(\tau)&:= \frac{\tau^2}{2\pi \nu} -\frac{\pi}{4} \label{eq: phi_nu} \\
  &\bar  \cE_{\tau|\alpha} = \mathcal{D}^\nu_{P_\nu,\Omega_\nu},&\quad
     P_\nu(\tau|\alpha) &:= \frac{1}{\sqrt{3 \pi \nu}} e^{-\frac{\left(\tau^{1/3}-2  (\pi  \nu )^{1/3} \alpha \right)^2}{2 (\pi  \nu )^{2/3}}}\label{eq: P_nu} \\
   && \Omega_\nu(\tau )&:= \frac{3(\pi \nu)^{\nicefrac{1}{3}}\tau^{\nicefrac{2}{3}}}{2} + \frac{\pi}{4}.\label{eq: omega_nu}
\end{align}

As in the large $\alpha$ limit the linear phase regime and the subsequent revivals are well separated in time, we can at all times write $\widetilde \cE_{\tau|\alpha}= \sum_{\nu=0}^\infty \mathcal{R}_{R_\nu,\Phi_\nu}^\nu $  and $\bar \cE_{\tau|\alpha}= \sum_{\nu=0}^\infty \mathcal{D}_{P_\nu,\Omega_\nu}^\nu$ which completes the derivation of the result.
\end{proof}

Let us now interpret the result in different time regimes and investigate the atomic QFI.

\subsubsection{Short times $\tau \ll \alpha$}
\label{sec: short t}

In the short time limit $\nicefrac{\tau}{\alpha}\to 0$, which was also analyzed in~\cite{PhysRevA.44.5913}, we find the non vanishing contributions to be $Q_0=\exp(-\nicefrac{\tau^2}{2})$ and $\Phi_0 =2\alpha \tau$, implying
\be \label{eq: short t}
\cE_{\tau|\alpha}[\bm \sigma] = 
\left(
    \begin{array}{c}
    \id \\
    e^{-\nicefrac{\tau^2}{2}} \big(\cos(2\alpha \tau) \sigma_x - \sin(2\alpha \tau)\sigma_z\big)
    \\
    \sigma_y 
    \\
    e^{-\nicefrac{\tau^2}{2}} \big( \sin(2\alpha \tau) \sigma_x + \cos(2\alpha \tau) \sigma_z \big)
    \end{array}
    \right).
\ee
This channel describes a rotation around the $y$-axis by the angle $2\alpha \tau$ (like in the semi-classical Rabi model) combined with the dephasing in the $y$-basis with parameter $e^{-\nicefrac{\tau^2}{2}}$ (due to the atom-field entanglement). Notice that the channel is not fully entanglement-breaking as long as the dephasing is not complete.\sout{, $r>0$.}

\begin{figure}
    \centering
    \includegraphics[width=0.98\columnwidth]{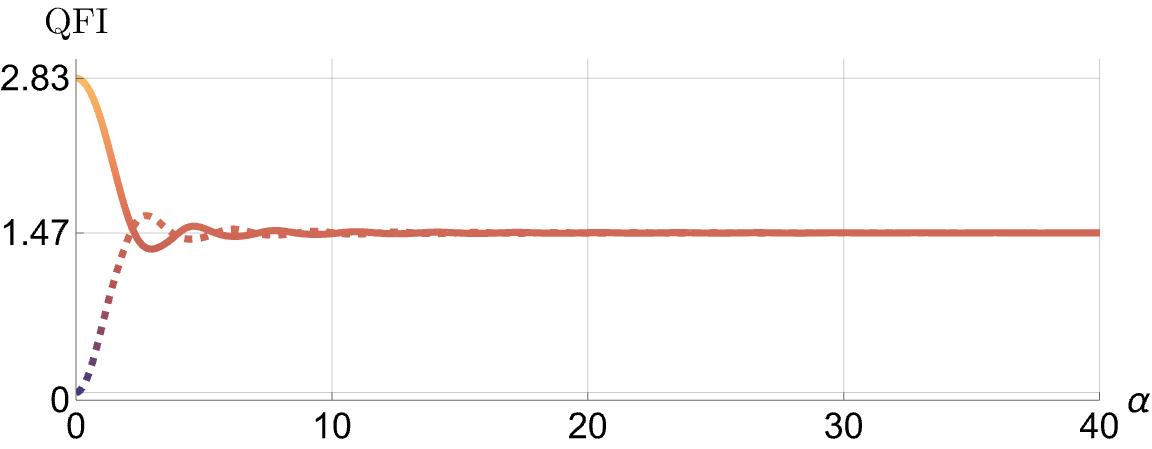}
    \caption{The ${\rm QFI}[\rho_{\tau=1|\alpha}^{(g/e)}]$ of the ground (full line) and excited (dashed line) initial state after an interaction with the coherent mode of duration $\tau=1$, as a function of $\alpha$. We see that it converges relatively fast to the asymptotic value $\nicefrac{4}{e}\approx 1.47$.}
    \label{fig:tau1}
\end{figure}

For the ground or excite initial state, one finds $x_{\tau|\alpha}^{(g/e)} = \pm e^{-\nicefrac{\tau^2}{2}} \sin(2\alpha \tau)$ and $z_{\tau|\alpha}^{(g/e)} = \pm e^{-\nicefrac{\tau^2}{2}} \cos(2\alpha \tau)$, which makes it easy 
to compute the QFI via Eq.~\eqref{eq: QFI xz}
\begin{equation} \label{eq: QFI short tau}
{\rm QFI}[\rho_{\tau|\alpha}^{(g/e)}]= 4\,  \tau^2 e^{-\tau^2}\leq 1.47.
\end{equation}
The upper-bound is saturated at time $\tau=1$ independent of $\alpha$, and yields a QFI which is about three times lower than the theoretical upper bound ${\rm QFI}[\rho_{\tau=1|\alpha}]\leq 4$. This shows that for a relatively short interaction time $\tau\approx 1$, a constant nonzero QFI is encoded in the atomic state for {\it all} values of $\alpha$. This is consistent with the numerical result in Figs.~\ref{fig: coh-pop} and \ref{fig: qfiground}, showing damped oscillation of the atomic population and coherence and a finite QFI which quickly decays. In Fig.~\ref{fig:tau1}, we plot the QFI for $\tau =1$, ground and excited initial states, and different values of $0\leq \alpha\leq 40$.

 To conclude the discussion of the fixed time limit, it is worth mentioning the ``textbook case'', where the effective interaction time $\tau = \frac{\theta}{2\alpha}$ decreases with $\alpha\to \infty$. The reduced dynamics of the atom becomes  $\cE_{\tau =\frac{\theta}{2\alpha}|\alpha}[\bullet] = e^{-\ii \frac{\theta}{2}\sigma_y} \bullet e^{\ii \frac{\theta}{2}\sigma_y}$ 
and describes a unitary Rabi-rotation by the desired angle. However, this regime is very suboptimal for the purpose of field amplitude estimation, with the QFI vanishing together with the interaction time.

\subsubsection{Long times $1 \ll \tau \ll \alpha^2$}

\label{sec: main limit}

When time increases $\tau \gg 1$ but remains relatively short $\tau \ll \alpha^2$, we find that the non-vanishing contributions are 
\be Q_{\nu}  =  \frac{1}{\sqrt{\pi \nu}}e^{-\left(\frac{\tau-2\pi\alpha\nu}{\sqrt 2\pi \nu}\right)^2}, \quad \Phi_{\nu} = \frac{\tau^2}{2\pi \nu} -\frac{\pi}{4},
\ee
with $\nu\geq 1$, and 
$P_0 = 1$ with $\Omega_0 = \frac{\tau}{2\alpha}$. As $Q_{\nu}(\tau |\alpha)$ decays exponentially with $(\nicefrac{\tau}{\pi \nu}-\alpha)^2$, the revivals are well separated in time, and the atomic evolution is given by 
\be\label{eq: channel longer}
\cE_{\tau|\alpha}[\bm \sigma] = 
\left(
    \begin{array}{c}
    \id +\sin(\nicefrac{\tau}{2\alpha})\sigma_x\\
    Q_\nu \big((-1)^{\nu} \cos (\Phi_{\nu}) \, \sigma_x - \sin(\Phi_{\nu})\, \sigma_z\big)
    \\
    \cos(\nicefrac{\tau}{2\alpha}) \sigma_y 
    \\
    Q_\nu \big( (-1)^{\nu}\sin(\Phi_{\nu}) \sigma_x + \cos(\Phi_{\nu}) \sigma_z \big)
    \end{array}
    \right)
\ee
where $\nu$ is to be chosen as the closest revival to $\tau$, i.e., minimizing $(\nicefrac{\tau}{\pi \nu}-\alpha)^2$. 
For the ground or excited initial condition, this gives \begin{align}
x_{\tau|\alpha}^{(g/e)}&=  \sin(\nicefrac{\tau}{2\alpha})\pm \,Q_\nu(\tau|\alpha) \,(-1)^\nu \sin( \frac{\tau^2}{2\pi \nu} -\frac{\pi}{4}),\\
 z_{\tau|\alpha}^{(g/e)}&= \pm \,Q_\nu(\tau|\alpha)\, \cos( \frac{\tau^2}{2\pi \nu} -\frac{\pi}{4} ).
\end{align}
This time regime can be further divided into two cases.\\

{\it At revivals $|\tau- 2\pi\alpha \nu| =O(1)$.} --- We find $\sin(\nicefrac{\tau}{2\alpha})=0$, $\cos(\nicefrac{\tau}{2\alpha})=(-1)^\nu$ in Eq.~\eqref{eq: channel longer}, hence the short-time dynamics is recovered up to an eventual phase. 
Such periodic revivals of rapidly changing atomic population $z_{\tau|\alpha}^{(g/e)}$ have been recognized in the literature~\cite{eberly1980periodic,phoenix1988fluctuations,gea1990collapse,buvzek1992schrodinger,shore1993jaynes,fleischhauer1993revivals}, and observed experimentally~\cite{rempe1987observation}. The revivals are very visible in the numerical results of Fig.~\ref{fig: coh-pop}  where the dashed lines correspond to $\tau = 2\pi \alpha$ and $\tau =4\pi \alpha$. Here, we see that the revivals also manifest in the atomic coherence $ x_{\tau|\alpha}^{(g/e)}$, and more generally allow one to recover some quantum information encoded in the original state. Due to the prefactor $\nicefrac{1}{\sqrt \nu}$, at longer times revivals disappear as $(\nicefrac{\tau}{\alpha})^{-\nicefrac{1}{2}}$. This is also the timescale at which the atomic population is scrambled $z_{\tau|\alpha}\to 0$ and the atomic dynamics becomes entanglement breaking, albeit for generic times (outside of revivals) this happens exponentially fast with $\tau^2$, as given by Eq.~\eqref{eq: short t}. Using Eq.~\eqref{eq: QFI xz}, close to revivals and for $R_\nu:= \frac{\tau-2\pi\alpha\nu}{\sqrt 2\pi \nu}$, the maximal QFI is found to be
\be\label{eq: QFI revival}
{\rm QFI}[\rho^{(g/e)}_{\tau|\alpha}] = \frac{8 R_\nu^2(\tau|\alpha) }{\pi  \nu \,  e^{2 R_\nu^2(\tau|\alpha)}-1}
\approx \frac{0.47}{\nu}
\ee
for $\tau \approx 2\pi \alpha \nu \pm \pi\nu$ (around the first revival $\nu=1$ the value is even slightly higher ${\rm QFI}(\rho_{\tau|\alpha}^{(g/e)})=0.54$, see  Appendix~\ref{app: large alpha limit}). This conclusion is in good agreement with the numerical results presented in Fig.~\ref{fig: qfiground}, where the dashed lines in the QFI plot correspond to $\tau = 2\pi \alpha \pm \pi$ and $\tau = 4\pi \alpha \pm 2\pi$, i.e., the optimal times around the first two revivals.\\

{\it Outside of revivals $|\tau -2\pi \nu \alpha| \gg 1$.} --- We find $Q_\nu \approx 0$ and the atomic dynamics is given by
\begin{align}
\cE_{\tau|\alpha}[\bm \sigma]=
     \left(
    \begin{array}{c}
    \id + \sin(\nicefrac{\tau}{2\alpha}) \sigma_x\\
   0
    \\
     \cos(\nicefrac{\tau}{2\alpha}) \sigma_y 
    \\
   0
    \end{array}
    \right), \label{eq: outside revivals}
\end{align}
describing complete dephasing in the $\sigma_y$ basis, followed by a generalized amplitude-damping channel. This channel is entanglement-breaking, since it involves a measurement in the $y$-basis.  For the atom prepared in the ground or excited states, the slow phase only contributes to the atomic coherence $x_{\tau|\alpha}^{(g/e)} =  \sin(\nicefrac{\tau}{2\alpha}),$ 
 $z_{\tau|\alpha}^{(g/e)}=0$. 
 These slow oscillations of the atomic coherence are well visible in the numerical simulation of Fig.~\ref{fig: coh-pop}. Particularly interesting are times $\tau \approx 2\pi \alpha (\nu +\nicefrac{1}{2})$ with integer $\nu$, where the atom is found in the pure state $\cE_{\tau|\alpha}[\rho_0]= \frac{1}{2}(\id + (-1)^\nu \sigma_x)$ independent of the initial condition. This can be seen in Fig.~\ref{fig: coh-pop}, where the blue and red lines correspond to $\tau = \pi \alpha $ and $\tau =3 \pi \alpha$, respectively. Here, the Jaynes-Cumming interaction acts as a SWAP gate --- the atomic state is prepared in a fixed pure state uncorrelated with the field, while, by unitarity, the latter encodes the initial state of the atom. This state-transfer effect has been discussed in Refs.~\cite{gea1990collapse,Zhang2015quantum}. Since the dynamics is governed by slow variations of the coherence with $\left|\frac{\dd}{\dd \alpha}\frac{\tau}{2\alpha}\right| = \frac{\tau}{2\alpha^2} =0$, we find that the quantum Fisher information is vanishing outside of revivals ${\rm QFI}(\rho_{\tau|\alpha}^{(g/e)})= 0$\footnote{Particular care must be taken around times $\tau \approx 2\pi \alpha (\nu +\nicefrac{1}{2})$ where the atomic state becomes pure. In this case, to show that ${\rm QFI}(\rho_{\tau|\alpha}^{(g/e)})= O(\alpha^{-2})$ we expand $x_{\tau|\alpha}^{(g/e)}$ to higher orders in $\alpha$.}.

\begin{figure*}
    \centering
    \includegraphics[width=\linewidth]{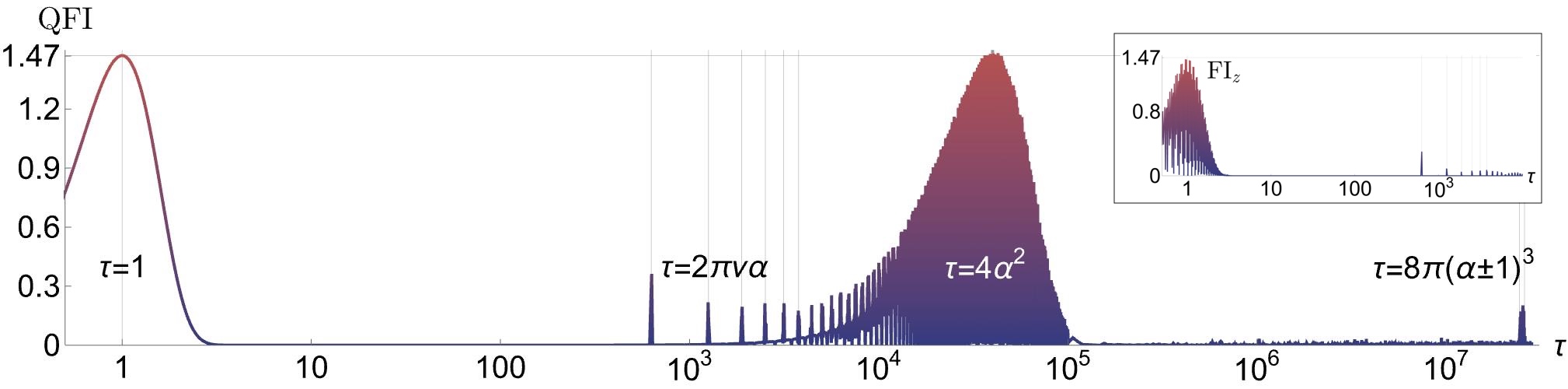}
    \caption{The QFI$(\rho_{\tau|\alpha}^{(g)})$ of the atom (initialized in the ground state) as a function of the interaction time for $\alpha =100$. The local maxima at $\tau =1$ and $4\alpha^2$, as well as the sharp revivals for $\tau =O(\alpha)$ and $O(\alpha^3)$, are predicted by the asymptotic analysis discussed in the text. The inset gives the Fisher information  ${\rm FI}_z(\rho_{\tau|\alpha}^{(g)})$ for the population measurement, which does not show the second maxima and the very late revivals.}
    \label{fig: new}
\end{figure*}

\subsubsection{Longer times $\alpha \ll \tau \ll \alpha^3$}

\label{sec: main limit}
For even longer times $\tau=O(\alpha^2)$, the revivals are no longer well separated; nevertheless, their contribution to the atomic dynamics is suppressed with $\nicefrac{1}{\sqrt \alpha}$. The atomic evolution is then found to be
\begin{align} 
\cE_{\tau|\alpha}[\bm \sigma]=
     \left(
    \begin{array}{c}
    \id + e^{-\nicefrac{\tau^2}{32 \alpha^4}} \sin(\nicefrac{\tau}{2\alpha}) \sigma_x\\
    O(\nicefrac{1}{\sqrt{\alpha}})
    \\
    e^{-\nicefrac{\tau^2}{32 \alpha^4}}\, \cos(\nicefrac{\tau}{2\alpha}) \sigma_y 
    \\
    O(\nicefrac{1}{\sqrt{\alpha}})
    \end{array}
    \right).\label{slow phase}
\end{align}
Remarkably, in this case the the quantum Fisher information is governed by $\left|\frac{\dd}{\dd \alpha}\frac{\tau}{2\alpha}\right| = \frac{\tau}{2\alpha^2} =O(1)$ and is nonzero. We find that it is rapidly oscillates between zero and 
\be\label{eq: QFI alphasq}
{\rm QFI}(\rho_{\tau|\alpha}^{(g/e)})= \frac{\tau^2}{4\alpha^4} e^{-\nicefrac{\tau^2}{16 \alpha^4}} \leq  1.47, 
\ee
attained at $\tau = 4 \alpha^2$, which reproduces the short time maximum of Eq.~\eqref{eq: QFI short tau} exactly. Due to the residual revival terms in Eq.~\eqref{slow phase}, this identity is only valid up to a correction of order $O(\nicefrac{1}{\sqrt \alpha})$, which is well visible in the Fig.~\ref{fig: new} where $\sqrt{\alpha}=10$ is still relatively low.

\subsubsection{Very long times  $ \tau \gg \alpha^2$}

\label{sec: long t}
At even longer times $\tau = O(\alpha^3)$, we find that due to the constructive interference of the slow phase, revivals of the atomic coherence also occur at much later times $\tau \approx 8 \pi \nu  \alpha ^3$
\begin{align}
\cE_{\tau|\alpha}[\bm \sigma]&=
     \left(
    \begin{array}{c}
    \id + (-1)^\nu P_\nu \sin(\Omega_\nu) \sigma_x\\
    0
    \\
     (-1)^\nu P_\nu \, \cos(\Omega_\nu) \sigma_y 
    \\
    0
    \end{array}
    \right),
\end{align}
with  
$P_\nu(\tau|\alpha) = \frac{1}{\sqrt{3 \pi \nu}} \exp(-\frac{\left(\tau^{1/3}-2  (\pi  \nu )^{1/3} \alpha \right)^2}{2 (\pi  \nu )^{2/3}})$ 
and $ \Omega_\nu(\tau )= \frac{3 (\pi \nu)^{\nicefrac{1}{3}}}{2}\tau^{\nicefrac{2}{3}} + \frac{\pi}{4}$.

This shows remarkable resilience of the atomic dynamics to chaos. While for generic times $\tau \gg \alpha$, the convergence of the atomic state to the maximally mixed one $\rho^{(g/e)}_{\tau|\alpha} \to \nicefrac{\id}{2}$ is governed by $e^{-\nicefrac{\tau^2}{32 \alpha^4}}$  in Eq.~\eqref{slow phase}. At specific times, with $P_\nu(\tau|\alpha)>0$ and $\cos\big(\Omega_\nu(\tau )\big)\approx 1$, the {\it classical} information encoded in the $y$-basis is partially recovered 
$\cE_{\tau|\alpha}[\frac{1}{2}(\id \pm \sigma_y)] = \frac{1}{2}(\id \pm P_\nu(\tau|\alpha) (-1)^\nu \sigma_y)$. Such revivals only disappear with $(\nicefrac{\tau}{\alpha^3})^{-\nicefrac{1}{2}}$. The QFI follows the same trend: it is rapidly oscillating close to revivals and reaches values up to
\be
{\rm QFI}[\rho^{(g/e)}_{\tau|\alpha}] \approx \frac{0.31}{\nu}
\ee
for $\tau \approx \pi  \nu  \left(2 \alpha \pm 2\right)^3$. This is illustrated in the numerical example of Fig.~\ref{fig: new}.

\subsubsection{Fully mixing times $\tau \gg \alpha^3$}

In the limit  $\tau \gg \alpha^3$ the atomic evolution is fully mixing
\be
\cE_{\tau|\alpha}[\rho_0]=
\mathcal{T}_{\nicefrac{\id}{2}}[\rho_0] = \nicefrac{\id}{2}.
\ee
We have also seen that it becomes entanglement-breaking at a shorter time-scale $\tau \gg \alpha$.

\subsubsection{Fisher information of the population measurement}

Saturating the QFI of an atomic state requires measuring the atom in a specific direction in the Bloch sphere. When this direction does not coincide with the $z$-axis, such a measurement needs a local oscillator in phase with the coherent mode of the field. It is therefore natural to also look at the Fisher information obtained from a passive measurement of the atomic population. This is equivalent to computing the QFI of a fully dephased state (setting the $x$ and $y$ components of the Bloch vector to zero), i.e.
\be
{\rm FI}_z(\rho_{\tau|\alpha}) = \frac{\dot z_{\tau|\alpha}^2}{1- z_{\tau|\alpha}^2}.
\ee

Revisiting the general expression of the QFI, we find that the Fisher information of the population measurement is rapidly oscillating together with the population. However, for well-selected times, it is equal to the QFI in the short times regime $\tau\ll \alpha$ (Eq.~\ref{eq: QFI short tau}) and at early revivals $\tau \approx 2 \pi \alpha \nu$ (Eq.~\ref{eq: QFI revival}). It vanishes at later times $\tau \gg \alpha$, where the atomic population is scrambled, $z\approx 0$. For $\alpha=100$, the Fisher information of the population measurement is plotted in the inset of Fig.~\ref{fig: new}.

\subsection{Beyond the Jaynes-Cummings encoding}

\label{sec: covariant interaction}

To conclude this section, let us come back to the general question --- what is the interaction encoding the real coherent states $\{\ket{\alpha}\}$ into a qubit probe with the globally optimal QFI (high for all $\alpha$) --- and briefly discuss another candidate.

As the real coherent states are aligned with the $\hat x =\frac{1}{\sqrt 2}(a+a^\dag)$ quadrature of the field, a very natural interaction is given by $\widetilde H = \frac{1}{2}\, \sigma_y (a+a^\dag)$ and
\be\label{eq: U tilde}
\widetilde U_\tau = e^{-\ii \tau \tilde H}= \int_{-\infty}^\infty \dd x \, e^{-\ii \frac{\tau x}{\sqrt 2} \sigma_y}\otimes \ketbra{x}
\ee
where $\ket{x}$ are the eigenstate of $\hat x$. It describes a controlled rotation 
\be\label{eq: mapping U}
\widetilde U: \ket{g}\ket{x} \mapsto \left(\cos(\frac{\tau x}{\sqrt 2})\ket{g} + \sin(\frac{\tau x}{\sqrt 2})\ket{e}\right)\ket{x}
\ee
of the probe by an angle proportional to the field quadrature $\hat x$. In other words, $\widetilde U$ realizes a von Neumann measurement of the field quadrature with the qubit pointer. For the real displacement operators $D(\alpha)= e^{\alpha(a - a^\dag)}$,  it has the natural covariance property $\widetilde U_\tau \, \mathcal{D}(\alpha) = \left(e^{-\ii \frac{\tau \alpha}{2}\sigma_y} \otimes \mathcal{D}(\alpha)\right) \,\widetilde U_\tau$, ensuring that $\widetilde U$ is equally suitable for all field amplitudes $\alpha$.

In this case, the reduced dynamics of the atom is immediately obtained, using $|\braket{x}{\alpha}|^2 = \frac{1}{\sqrt{\pi}} \exp(-(x -\sqrt{2} \alpha)^2)$, one finds 
\begin{align}
\widetilde \cE_{\tau|\alpha}[\bullet] &=\tr_{\rm field} \widetilde U_\tau (\bullet \otimes \ketbra{\alpha} )\widetilde U_\tau^\dag \\
&= \frac{1}{\sqrt \pi}\int \dd x \, e^{-(x -\sqrt{2} \alpha)^2}\, 
\mathcal{Y}_{\frac{\tau x}{\sqrt{2}}} [\bullet]
\end{align} 
where $\mathcal{Y}_\varphi [\bullet] = e^{-\ii \varphi \sigma_y} \bullet e^{\ii \varphi \sigma_y}$ is the unitary rotation around $\sigma_y$. Performing the Gaussian integral gives
\be
\widetilde \cE_{\tau |\alpha} [\bm \sigma] = 
    \left(\begin{array}{c}
    \id \\
    e^{-\nicefrac{\tau^2}{2}} \big(\cos(2\alpha \tau) \sigma_x - \sin(2\alpha \tau)\sigma_z\big)
    \\
    \sigma_y 
    \\
    e^{-\nicefrac{\tau^2}{2}} \big( \sin(2\alpha \tau) \sigma_x + \cos(2\alpha \tau) \sigma_z \big)
    \end{array}\right).
\ee
Remarkably, we recognize here the dynamics induced by the Jaynes-Cummings interaction (Eq.~\ref{eq: short t}) in the limit of short time $\tau\ll \alpha$ and large amplitude $\alpha\gg 1$. Hence, the atomic ${\rm QFI}[\widetilde \rho_{\tau|\alpha}^{(g/e)}] \leq 1.47$ is given by the same expression maximized at $\tau=1$.


\section{Interactions of an atom with a source of coherent radiation}
\label{sec: sequential}

As we have seen, regardless of the amplitude of the coherent field, the model where a single mode interacts with the atom is very different from an atom interacting with a classical field of unknown amplitude.  A natural generalization that often better describes the driving of an atom by a harmonic field (as e.g. emitted by a laser) is its interaction with a \emph{source} of coherent radiation. We discuss this situation in the context of a simple collision model, where the atom sequentially interacts with a series of coherent modes $\ket{\alpha}$  with unknown amplitude, and the interaction is again described by the resonant Jaynes-Cummings model.

\subsection{A sequence of $N$ interactions}

First, consider the situation where the atom interacts with a sequence of a fixed number $N$ of coherent modes. Each interaction has a duration $\tau= g t$ resulting in the channel $\cE_{\tau|\alpha}$ in Eq. \eqref{eq: channel in Gramm}.  After the $N$ steps, the reduced evolution of the atom is thus governed by 
\begin{equation}
  \cE^{\circ N}_{\tau|\alpha} =\underbrace{\cE_{\tau |\alpha}\circ \dots \circ \cE_{\tau |\alpha} }_{N}.
\end{equation}
We are interested in the final state of the atom $\rho_{\tau,N|\alpha} :=  \cE_{\tau |\alpha}^{\circ N}[\Psi_0]$ and the maximal ${\rm QFI}[\rho_{\tau,N|\alpha}]$ 
that can be obtained after $N$ interactions.\\

In the limit $\alpha\gg 1$, we can use the general expression of the atomic evolution, computed in Sec.~\ref{sec: main limit}, to derive the optimal quantum Fisher information. 

First, consider the regime where the time $\tau$ is short (Eq.~\ref{eq: short t}) or close to revivals  (Eq.~\ref{eq: channel longer} with $\cos(\nicefrac{\tau}{2\alpha})=1$). In both cases, the atomic evolution is given by a rotation around $y$ by the angle $\Phi_\nu$ composed with a dephasing in the $y$ basis of strength $Q_\nu$, plus a potential phase flip $\sigma_{x(y)} \to (-1)^\nu \sigma_{x(y)}$. Composing such channels $N$ times, we find 
\be
     \mathcal{E}_{\tau|\alpha}^{\circ N} :=\left(
    \begin{array}{c}
    \id\\
    Q_\nu^{N} \big( \cos(N \Phi_\nu) \sigma_x - \sin(N \Phi_\nu)\sigma_z\big)
    \\
   \sigma_y
    \\
     Q_\nu^{N} \big( \sin(N \Phi_\nu) \sigma_x + \cos(N \Phi_\nu)\sigma_z\big)
    \end{array}
    \right)
\ee
for even $\nu$. While for odd $\nu$, subsequent rotations cancel each other because of the phase flip, so the rotation angle is $\Phi_\nu$ or $0$ depending on the parity of $N$.

 For $\nu = 0$, it is only the phase $\Phi_\nu = 2 \alpha \tau$ that depends on $\alpha$, while $Q_0=e^{-\nicefrac{\tau^2}{2}}$. For the QFI of any real initial state (which is optimal), we find
\be\label{eq: QFI N times}
{\rm QFI} = 4 N^2 \tau ^2 \, e^{-N \tau ^2}.
\ee

By comparing the expressions for $N=1$ and $2$, we conclude that it is only beneficial to increase the number of interactions beyond one if $\tau\leq \sqrt{\log 4}\approx 1.18$. 

At revivals $\nu\geq 1$, it is only the dephasing rate $Q_\nu= e^{-\left(\frac{\tau-2\pi\alpha\nu}{\sqrt 2\pi \nu}\right)^2}/\sqrt{\pi \nu}$ which depends on $\alpha$, and the QFI for any real initial state is given by 
\begin{align}
{\rm QFI} = \frac{8 N^2 R_\nu^2}{(\pi \nu)^N e^{2 N R_\nu^2}-1} &\leq -4 N\,  W_0\left(-\frac{1}{e (\pi \nu)^N}\right)\\
&\approx \frac{4 N}{e (\pi  \nu )^N},
\end{align}
where we optimized over $R_\nu=\frac{\tau-2\pi\alpha\nu}{\sqrt 2\pi \nu}$ to get the inequality. Here, we find that the optimal integer value of $N$ is, in fact, always $N=1$, so applying the channel more than once never increases the QFI.

In turn, for times $1\ll \tau\ll \alpha^2$ and outside of the revivals, the atomic evolution composed $N$ times reads (see Eq.~\eqref{eq: outside revivals})
\be
     \mathcal{E}_{\tau|\alpha}^{\circ N} :=\left(
    \begin{array}{c}
    \id + \sin(\nicefrac{\tau}{2\alpha}) \sigma_x \\
   0
    \\
   \cos^{N}(\nicefrac{\tau}{2\alpha}) \sigma_y
    \\
   0
    \end{array}
    \right).
\ee
Here again, more than one channel application can never improve the QFI. In fact, one can verify that this conclusion also remains true for longer interaction times $\tau$.\\

We conclude that it is only in the short time limit $\tau\leq 1.18$ (or $t \leq \nicefrac{1.18}{g}$), where the subsequent interactions are beneficial for field estimation. We can then optimize the expression of the QFI in Eq.~\eqref{eq: QFI N times} with respect to the individual interaction time $\tau$ to obtain 
\be\label{eq: QFI cuss JC}
\max_{\tau } {\rm QFI}[\rho_{\tau,N|\alpha} ] = \frac{4}{e} N \approx 1.47 N 
\ee
for $\tau =\nicefrac{1}{\sqrt{N}}$ (or $t= \nicefrac{1}{g\sqrt{N}}$) independent of $\alpha$. 
This can be compared to the theoretical upper-bound implied by monotonicity. The QFI of the final atomic state can not exceed that of the initial states $\ket{\alpha}^{\otimes N}$ of the radiation modes, which is additive for a product state and equals $4N$, i.e. 
\be\label{eq: QFI series UB}
{\rm QFI}[\rho_{\tau,N|\alpha} ] \leq 4 N.
\ee
Again, we find that, also in the $N$ interactions scenario, the Jaynes-Cummings model is only a factor $e$ lower than the theoretical maximum. It is natural to ask whether the upper-bound $4 N$ is saturable in the sequential scenario with a two-level probe. We show in the App. \ref{app: optimal succ int} that this is indeed the case; however, similarly to $V$ in Eq.~\eqref{eq: V opt}, the derived interactions are fine-tuned to the value of $\alpha$ and moreover change between successive interaction steps. It is an interesting open question whether the QFI in Eq.~\eqref{eq: QFI cuss JC} can be exceeded with an interaction suitable for a large range of $\alpha$ and/or time-independent.
\\

Next, we discuss the scaling of the QFI with respect to the total interaction time $T=N \tau$. Eq.~\eqref{eq: QFI series UB} ${\rm QFI}[\rho_{\tau, N|\alpha} ] \leq 4 \frac{T}{\tau}$  implies that for any \emph{fixed} $\tau$, QFI is bound to a linear scaling with total time. With this in mind we examine the limit $\tau\to 0$, where the QFI could a priori scale super-linearly with $T$. We consider two different ways to take the limit, depending on whether the total energy of the coherent modes is allowed to diverge or must be kept bounded.

\subsection{The infinite modes limit}

First, we explore the limit where the single interaction time $\tau = g\, \dd t \to 0$ is small while $N=\frac{t}{\dd t}\to \infty$ and the amplitude of the coherent states $\alpha$ remains constant. 
To recover the dynamics of the atom in the  limit, we expand the states in Eq.~\eqref{eq: state gens} in small $\tau$
 \begin{equation}\label{eq: state gens coll}
 \begin{split}
\ket{\alpha_0} &=\left( 1-\frac{\tau^2 \hat n}{2}\right) \ket{\alpha},\quad \qquad  \,\,\,\ket{\alpha_2} =-\tau a^\dag \ket{\alpha}, 
\\
\ket{\alpha_1} &=\left(1-\frac{\tau^2 (\hat n+1)}{2}\right)\ket{\alpha},\quad
\ket{\alpha_3} = \tau a \ket{\alpha}.
\end{split}
\end{equation}
Tracing out the field mode, we find (see App.~\ref{app: infinite modes} for details) that the channel is unitary $
    \cE_{\dd t|\alpha}[\bullet]=\bullet -\ii \, \alpha \, g\, \dd t [\sigma_y, \bullet] +O(\dd t^2)$, and the atomic evolution is dictated by the master equation  
\begin{equation}
\frac{\dd }{\dd t} \rho_{t|\alpha} = - \ii g \, \alpha \,  [\sigma_y,\rho_{t|\alpha} ],
    \label{eq: seq-t-small}
\end{equation}
exactly reproducing the semi-classical Rabi model of Eq.~\eqref{eq: H class} for $\alpha =E$ with ${\rm QFI}=4 \, g^2 t^2$. Furthermore, for large $\alpha\gg 1$ this can be seen to be the maximal atomic QFI for the Jaynes-Cummings interaction with an unlimited number of modes and total time $t$. Indeed, plugging $\tau = g\,\dd t $ and $N= \frac{ t}{\dd t}$ in Eq.~\eqref{eq: QFI N times} we find
\be
{\rm QFI} = 4\, g^2 t^2 e^{-g^2 t\dd t} \leq  4 \, g^2 t^2
\ee
saturated in the limit $\dd t \to 0$.  The infinite modes limit is an example of the ideal Lamb-shift metrology theorized in \cite{sekatski2022optimal}: the coupling of the atomic probe to the field modes (the sample) is fully captured by a parametric Hamiltonian term (Lamb-shift) in Eq.~\eqref{eq: seq-t-small} and does not introduce any dissipation.\\

However, this limit is arguably nonphysical, as it corresponds to a diverging energy of the field modes $N\alpha^2\to \infty$, and diverging photon flux $\frac{\alpha^2}{\dd t} \to \infty$  (also diverging total QFI of the coherent modes $4 N \to \infty$). For this reason, we now consider an alternative limit, where these quantities remain bounded.

\subsection{The continuous field limit}

To obtain a physical limit with a bounded energy we take $\alpha^2 = \varepsilon^2 \dd t$, such that $\varepsilon^2$ is the photon flux, and the amplitude $\varepsilon$ is our parameter of interest.  To retain a non-vanishing interaction we then set the interaction parameter scaling to $\tau = \sqrt{\kappa \,  \dd t}$, as common in quantum optics~\cite{gardiner2004quantum,breuer2002theory,ciccarello2022quantum}. In the Appendix~\ref{app:collision}, we show that in this limit, one recovers the model with the atom coupled to a one-dimensional bosonic field with the time-dependent Hamiltonian
\be
H(t) = \ii\,  \sqrt{\kappa} \Big(\sigma_- (\varepsilon + a^\dag(t)) - \sigma_+ (\varepsilon+a(t))\Big),
\ee
where the bosonic field operators obey the usual commutation relations $[a(t),a^\dag (t' )]= \delta(t-t')$ and have vacuum statistics. See e.g. Refs.~\cite{Fischer2018scatteringintoone,dkabrowska2021eternally} where this model is discussed. 

To obtain the reduced dynamics of the atom in the continuous field limit, we plug $\tau =  \sqrt{\kappa \dd t}$ and $\alpha = \varepsilon \sqrt{\dd t}$ in Eq.~\eqref{eq: state gens coll}, giving the infinitesimal channel 
\be
\cE_{\dd t|\varepsilon}[\bullet] = \bullet + \dd t\left(- \ii \sqrt{\kappa} \,\varepsilon [\sigma_y,\bullet] + \kappa \, \mathcal{L}_{\sigma_-}[\bullet]\right) + O(\dd t^2), 
\ee
  where $\mathcal{L}_{A}[\bullet] = A \bullet A^\dag- \frac{1}{2} \{A^\dag A , \bullet \}$ (see Appendix~\ref{app:collision} for details). Rearranging the terms, one concludes that the atomic dynamics is governed by the master equation
 \be\label{eq: master 1}
 \frac{\dd }{\dd t} \rho_{t|\varepsilon} = - \ii \sqrt{\kappa} \,\varepsilon \,  [\sigma_y,\rho_{t|\varepsilon} ] + \kappa \, \mathcal{L}_{\sigma_-}[\rho_{t|\varepsilon}],
 \ee
 where one recognizes the semi-classical Rabi model with the additional spontaneous emission term given by the dissipator $\mathcal{L}_{\sigma_-}$. 
The calculation we just did is not new, and the conclusion is even older~\cite{Einstein} -- coupling an atom to quantized radiation enables spontaneous emission. For our task of sensing $\varepsilon$ with the atomic probe, this implies that its QFI can not increase indefinitely. \\

  To solve the master equation~\eqref{eq: master 1}, it is convenient to work in time units $s=\kappa t$ dictated by the coupling constant and introduce the rescaled field amplitude $\bar \varepsilon = \frac{\varepsilon}{\sqrt{\kappa}}$. With this variable change the master equation simply reads
\be\label{eq: master}
 \frac{\dd }{\dd s} \rho_{s|\bar \varepsilon} = - \ii  \,\bar \varepsilon \,  [\sigma_y,\rho_{s|\bar  \varepsilon} ] +  \mathcal{L}_{\sigma_-}[\rho_{s|\bar  \varepsilon}].
 \ee
 \begin{figure*}[t]
    \centering
    \includegraphics[width=0.49\linewidth]{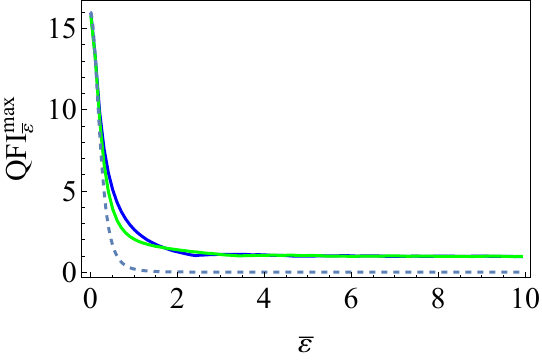}
     \includegraphics[width=0.49\linewidth]{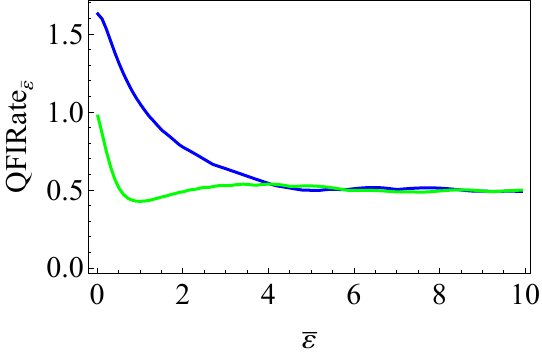}
    \caption{{\bf (Left)} The optimal QFI  as a function of $\bar\varepsilon$  for the initial ground state (blue) and for the initial excited state (green). The dashed line corresponds to the steady QFI computed in Eq.~\eqref{eq: steady-QFI}. {\bf (Right)} The optimal QFI rate as a function of $\bar\varepsilon$  for the initial ground state (blue) and for the initial excited state (green). The theoretical upper bound of 4 (also valid for adaptive processes with unlimited auxiliary entanglement) follows from Refs.~\cite{sekatski2017quantum, demkowicz2017adaptive,zhou2018achieving}.}
    \label{fig: opt-qfi}
\end{figure*}
 This equation~\eqref{eq: master} is straightforward to solve analytically; however, the expressions one obtains are not very insightful. Instead, we now focus on its consequences for the field estimation task. 

 \subsubsection{The optimal QFI}
 
As the first quantity of interest, we discuss the maximal QFI reached by the atomic state (at some optimal time $s$)
\be
{\rm QFI}^{\max}_{\bar \varepsilon} = \sup_s {\rm QFI}[\rho_{s|\bar \varepsilon}].
\ee
The QFI of the state $\rho_{s|\bar \varepsilon}$ can also be computed analytically.  It admits a particularly simple expression at $\bar \varepsilon =0$,  where for the ground and excited initial states one finds (see App.~\ref{app: QFI-epsilon-bar} for an arbitrary real initial state)
\begin{align}\label{eq: QFI ground cont}
{\rm QFI}[\rho^{(g)}_{s|\bar \varepsilon=0}]&=16 \, e^{-s/2} \left(e^{s/4}-1\right)^2,\\
{\rm QFI}[\rho^{(e)}_{s|\bar \varepsilon=0}]&=16 \,e^{-s} \left(e^{s/2}-3\, e^{s/4} +2\right)^2.
\label{eq: QFI excited cont}
\end{align}

The dependence of the maximal QFI on $\bar \varepsilon$ is plotted in Fig.~\ref{fig: opt-qfi} (left panel) for the ground (blue) and excited (green) initial states. We see that it is maximal at zero field $\bar \varepsilon=0$, where $\lim_{s\to \infty}{\rm QFI}[\rho^{(g/e)}_{s|\bar \varepsilon=0}]=16$ follows from Eqs.~(\ref{eq: QFI ground cont}, \ref{eq: QFI excited cont}). While for $\varepsilon\gg 1$ it saturates the value ${\rm QFI}^{\rm max}_{\infty}\approx 0.97$, for which the optimal interaction time $s \approx 2.52$ is again independent of $\bar \varepsilon$.

\subsubsection{The steady-state QFI}

Next, let us consider the steady state $\rho^*_{\bar  \varepsilon}$ of the dynamics. Solving $0= - \ii  \,\bar \varepsilon \,  [\sigma_y,\rho^*_{\bar  \varepsilon} ] +  \mathcal{L}_{\sigma_-}[\rho^*_{\bar  \varepsilon}]$ we find the atomic steady-state read 
 \be
\rho^*_{\bar  \varepsilon} = \frac{1}{2}\left( \id +\frac{1}{1+8 \bar \varepsilon^2}\, \sigma_z + \frac{4 \bar \varepsilon}{1+8 \bar \varepsilon^2} \,\sigma_x\right),
 \ee
 and gives 
\be \label{eq: steady-QFI}
{\rm QFI}[\rho_{\bar \varepsilon}^*] = \left(\frac{4}{(1+8 \, \bar \varepsilon^2)}\right)^2.
\ee
The QFI of the steady state is equal to $16$ at $\bar \varepsilon=0$ and vanishes in the limit of large $ \bar \varepsilon$, where the steady state is maximally mixed. This is in sharp contrast with the optimal transient ${\rm QFI}^{\max}_{\bar \varepsilon}$, which is finite for all  $\bar \varepsilon$.

\subsubsection{The optimal QFI rate}

In the limit where the total interaction time is very long, the best possible strategy (from the Fisher information perspective) is to repeat the measurement procedure for which the QFI rate is maximized 
\be
{\rm QFIRate}_{\bar \varepsilon}:=\sup_s \frac{{\rm QFI}[\rho_{s|\bar \varepsilon}] }{s},
\ee
see e.g. Refs.~\cite{correa2015individual,mehboudi2019thermometry,sekatski2017quantum}. 

In Fig.~\ref{fig: opt-qfi} (right panel), the maximal QFI rate as a function of $\bar{\varepsilon}$ for the initial ground and excited states. For the initial ground state, the behavior is simple since the ${\rm QFIRate}_{\bar \varepsilon}$ is monotonically decreasing w.r.t. $\bar{\varepsilon}$. For $\bar \varepsilon=0$ we can compute the optimal rate using Eq.~\eqref{eq: QFI ground cont}, we find that  
\be
{\rm QFIRate}_{\bar \varepsilon=0}^{(g)} \approx 1.63
\ee
is attained at $s\approx 5.03$.\\

\subsubsection{The impact of $\kappa$ and summary}

Finally, let us interpret these results in terms of the physical time $t= \frac{s}{\kappa}$ and the physical amplitude $\varepsilon = \sqrt{\kappa} \bar \varepsilon$. For a state $\rho_\varepsilon$ we denote the QFI with respect to the $\varepsilon$ as $\widetilde{\rm QFI}[\rho_{\varepsilon}]$. To relate it to the quantities computed above, note that the QFI transforms as a metric (see Eq.~\ref{eq: qfi bures})
\begin{equation}\label{eq: reparam}
\widetilde{\rm QFI}[\rho_{\varepsilon}]=\left(\frac{\dd \bar \varepsilon}{\dd \varepsilon}\right)^2 {\rm QFI}[\rho_{\bar \varepsilon}]=\frac{1}{ \kappa} {\rm QFI}[\rho_{\bar \varepsilon}].
\end{equation}
For a single interaction, the maximal reachable QFI, expressed in the physical amplitude $\varepsilon$, can thus be increased when the interaction parameter $\kappa$ is lowered. This can be understood from the fact that lowering $\kappa$ delays the onset of decay in Eq.~\eqref{eq: master 1}, relatively to the coherent term, effectively prolonging the coherent sensing time. 
For $\varepsilon\neq 0$, in the limit of small $\kappa \ll \frac{1}{\varepsilon^2}$ the maximal QFI becomes
\be
\widetilde {\rm QFI}^{\max}_{\varepsilon} =  \frac{1}{\kappa}{\rm QFI}^{\max}_{\varepsilon/\sqrt \kappa} \approx \frac{{\rm QFI}^{\max}_{\infty}}{\kappa } \approx \frac{0.97}{\kappa}.
\ee
The optimal transient QFI is thus diverging for all $\varepsilon$ in the limit of vanishing $\kappa$. However, the price to pay is the increase of the interaction time $t \sim \nicefrac{1}{\kappa}$ required to reach the optimal atomic state.

In turn, when looking at the optimal QFI rate expressed in terms of the physical quantities, we do not find any divergencies
\begin{align}\label{eq: rate final}
\widetilde{\rm QFIRate}_{\varepsilon}&=\sup_t \frac{\widetilde{\rm QFI}[\rho_{t| \varepsilon}] }{t} = \sup_s \frac{{\rm QFI}[\rho_{s| \varepsilon/\sqrt{\kappa}}] }{\kappa \cdot \nicefrac{s}{\kappa}} \\
&= \,  {\rm QFIRate}_{ \varepsilon/\sqrt{\kappa}} \leq 1.63,
\end{align}
with optimal interaction time $t \sim \nicefrac{1}{\kappa}$ again decreasing with the interaction strength. For the QFI rate, increasing $\kappa$ is typically beneficial since ${\rm QFIRate}_{ \bar \varepsilon}$ can reach higher values when $ \bar \varepsilon$ approaches zero, as can be see from Fig.~\ref{fig: opt-qfi}.

The bound~ \eqref{eq: rate final} is to be compared with the QFI rate (or flux) emitted by the radiation source itself
\be
\frac{\widetilde {\rm QFI}[\rho_{t| \varepsilon}]}{t} \leq \frac{N \, {\rm QFI}\big(\ket{\alpha_{ \varepsilon}}\big)}{t} \to \frac{{\rm QFI}\left(\ket{\epsilon \sqrt{\dd t}}\right)}{\dd t} = 4.
\ee 
where we used $N = \frac{t}{\dd t} = \frac{s}{\kappa \dd t}$ and $\alpha = \varepsilon \sqrt{\dd t}$ and the parameter change formula \eqref{eq: reparam}. Interestingly, he same upper-bound ${\rm QFIRate}_{\bar \varepsilon}\leq 4$ can be derived by applying the general techniques of~\cite{sekatski2017quantum, demkowicz2017adaptive,zhou2018achieving} to the master equation~\eqref{eq: master}. Note that approaching this bound in general requires entangling the atom with an auxiliary system and using fast adaptive control. Furthermore, this entanglement and control are known to be helpful in the case of spontaneous emission noise, as they enable error-detection techniques. It is thus natural that our simple sequential probing of the field does not saturate the theoretical maximum.



The optimal choice of $\kappa$ thus depends on the considered scenario, for instance, whether the total interaction time or the total number of shots (e.g., limited by long re-initialization time) is the limiting resource. Generally however, we see that the QFI rate emitted by the radiation source is constant. Hence, for long times no sensing strategy can yield a QFI increasing quadratically in time, moreover its quadratic scaling at short times is synonymous of a suboptimal interaction.



\section{Conclusion}

In this paper, we studied the task of estimating the amplitude of a coherent quantum field with a two-level atomic probe, as modeled by the resonant Jaynes-Cummings interaction. We have considered two scenarios -- where the atom interacts with a single coherent mode of the electromagnetic field, or with a sequence of such modes (modeling a source of coherent radiation). Focusing on the probe's QFI to quantify the estimation precision, we have seen that in both scenarios its dependence on the field amplitude and the interaction time $\tau$ is very different from the semi-classical Rabi model, where the field amplitude enters as a scalar parameter of the Hamiltonian and ${\rm QFI} = 4 \tau^2$ for all field amplitudes.  

In the single-mode scenario (e.g. a single cavity mode), the QFI is bounded by 4, a constant dictated by the non-orthogonality of the coherent states $\ket{\alpha}$ themselves, and can not increase indefinitely with the interaction time. In addition, approaching this bound with the atomic probe requires an interaction that faithfully encodes the subspace $\{\ket{\alpha},\ket{\dot \alpha}\}$ into the two-level atom. For the Jaynes-Cummings interaction, we found that this is only realized in the vacuum limit $\alpha \ll 1$, where the QFI periodically approaches the maximal value of 4 for both ground and excited initial states of the atom. In the opposite limit $\alpha\gg 1$, the reduced evolution of the atom can also be solved analytically, see Eq. \eqref{eq: channel general} of Sec.~\ref{sec: main limit}, a result  that might be of independent interest in quantum optics. Here, we found that the maximal ${\rm QFI}\approx1.47$ is attained at $\tau=1$ and $\tau = 4 \alpha^2$. The first maximum is attained at a fixed time (independent of $\alpha$) and is conjectured to realize the optimal encoding of the real coherent states into a qubit probe. The second maximum requires measuring the atomic coherence and is absent for the population measurement.

In the sequential scenario (e.g. a chiral waveguide), the atom interacts with a sequence of $N$ coherent modes $\ket{\alpha}$ for a total time $T=N \tau$. In the large $\alpha\gg 1$ limit, using the analytic expression of the reduced dynamics, we found 
that the optimal ${\rm QFI} \approx 1.47 \, N$ is achieved for the interaction times $\tau =\nicefrac{1}{\sqrt N}$ independent of $\alpha$, and is close to the theoretical limit ${\rm QFI} \leq 4N$. Motivated by this bound, we then studied the continuous limit $N\to \infty$ and $\tau \to 0$, which can be taken in two ways. First, for $\tau = g\, \dd t$, $N=\frac{t}{\dd t}$ and constant $\alpha$ one recovers the semi-classical Rabi model with $E=\alpha$. However, this limit corresponds to a diverging photon-flux $\nicefrac{\alpha^2}{\dd t}\to \infty$, and is ultimately nonphysical. Second, for
$\tau= \sqrt{\kappa \, \dd t}, \alpha = \varepsilon \sqrt{\dd t}$ and $N =\nicefrac{1}{\dd t}$, one recovers the model where the atom interacts with a continuous one-dimensional field and $\varepsilon$ becomes the parameter of interest. Here, the atom follows a simple Lindbladian evolution that resembles the semiclassical Rabi model with an important caveat: it is driven by the Rabi frequency $\sqrt{\kappa} \,\varepsilon$, but it is also subject to spontaneous emission noise with rate $\kappa$, which forbids the QFI from increasing indefinitely. This is a direct consequence of the fact that the QFI flux produced by such radiation source is a bounded constant ${\rm QFI}\left(\ket{\epsilon \sqrt{\dd t}}\right)/ \dd t = 4$. Hence, in the long time limit all sensing strategies are bound to a linear scaling of the QFI with time. A quadratic scaling with time is only sustainable for short times and is synonymous of a suboptimal interaction\footnote{Taking the derivation of the App.~\ref{app: optimal succ int} to the continuous limit, shows that there always exists an $\epsilon$ and time-dependent interaction which maps all the QFI emitted by the radiation source on a two-level probe.}


Our results show that the semi-classical Rabi model is only a good approximation for the quantum-field estimation task in a specific limit, which is ultimately nonphysical. Outside this limit, we found that the QFI of the atomic probe behaves very differently from the semi-classical prediction: 
In the single-mode scenario it is limited by the non-orthogonality of coherent states. In the continuous-field scenario it is limited by a constant QFI flux produced by the source, and by the spontaneous emission noise unavoidably present in the reduced dynamics of the atomic probe due to back-action on the field. The complementarity of these arguments gives a fresh ``informational'' insight on the origin of spontaneous emission in the interaction of quantum radiation with matter~\cite{Einstein,simon2000optimal}.

Throughout the paper, we have identified several open questions regarding the optimal encoding of the QFI carried by large quantum systems (the field modes) into a small quantum probe (the two-level atom). We believe that many more such questions remain to be uncovered in the topic of quantum estimation of ``quantum parameters''.

\acknowledgements
R.R.R. is thankful to Konrad Schlichtholz and Javier Rivera-Dean for useful comments. P.S. thanks Mikael Afzelius and Riccardo Castellano for helpful and stimulating discussions. R.R.R. is financially supported by the Ministry for Digital Transformation and of Civil Service of the Spanish Government through the QUANTUM ENIA project call - Quantum Spain project, and by the EU through the RTRP - NextGenerationEU within the framework of the Digital Spain 2026 Agenda, and by Conselleria d’Educació i Universitats del Govern de les Illes Balears and Fons Social Europeu+ through the contract POSTDOC2024\_17.  M.P.-L. acknowledges support from the Grant ATR2024-154621 funded by MICIU/AEI/10.13039/501100011033. P.S. acknowledges support from the Swiss National Science Foundation via NCCR-SwissMap and from the Swiss Secretariat for Education, Research and Innovation (SERI) under contract number UeM019-3.

\bibliographystyle{bibstyle}
\bibliography{bibfile.bib}
\onecolumngrid
\newpage
\appendix

\section{QFI of a two-dimensional system}

\label{app:QFI-general}

For a general two dimensional state $\rho_\theta = \left(\begin{array}{cc}
 a_\theta & b_\theta-\ii c_\theta \\
 b_\theta + \ii c_\theta & 1-a_\theta 
\end{array}\right)$, and its derivative $\dot \rho_\theta \equiv \frac{\partial \rho_\theta}{\partial \theta}= \left(\begin{array}{cc}
 \dot a_\theta & \dot b_\theta-\ii \dot c_\theta \\
 \dot b_\theta + \ii \dot c_\theta & -\dot a_\theta 
\end{array}\right)$ one can compute the SLD $L$ by means of the Lyapunov equation
\be
\dot \rho_\theta = \frac{L \rho + \rho L}{2},
\ee
and obtain
\be
L= \left(
\begin{array}{cc}
 \frac{\dot a \left(a+2 b^2+2 c^2-1\right)-2 (a-1) (b \dot b +c \dot c)}{(a-1) a+b^2+c^2} & \frac{-((2 a-1) \dot a  (b-i c))+2 \dot b \left((a-1) a+i b c+c^2\right)-2 i \dot c ((a-1) a+b (b-i c))}{(a-1) a+b^2+c^2} \\
 \frac{-((2 a-1) \dot a (b+i c))+2 \dot b \left((a-1) a-i b c+c^2\right)+2 i \dot c ((a-1) a+b (b+i c))}{(a-1) a+b^2+c^2} & \frac{a (\dot a+2 b \dot b+2 c \dot c)-2 \dot a \left(b^2+c^2\right)}{(a-1) a+b^2+c^2} \\
\end{array}
\right),
\ee
where we removed the dependence of $a$, $b$ and $c$ on $\theta$ to avoid the cluttering. With the SLD, we can directly compute the QFI and obtain
\be
\label{eq: QFI-2x2}
{\rm QFI} (\rho_\theta) = \tr (L^2 \rho_\theta) = \frac{4 \left(a^2 \left(\dot b^2+\dot c^2\right)-a \left(\dot b^2+\dot c^2\right)+(\dot b c-b \dot c)^2\right)-4 (2 a-1) \dot a (b \dot b+c \dot c)+\dot a^2 \left(4 b^2+4 c^2-1\right)}{(a-1) a+b^2+c^2}.
\ee 
Noting that $a=(z+1)/2$, $b=x/2$, and $c=y/2$, where $x$, $y$, and $z$ are the usual Bloch vector components, we can rewrite Eq.~\eqref{eq: QFI-2x2} as  
\begin{equation}
    {\rm QFI}(\rho_\theta)=\frac{\|\dot{\bm r}_\theta\|^2 - \|\dot{\bm r}_\theta \cross \bm r_\theta\|^2}{1-\|\bm r_\theta\|^2 },
\end{equation}
where $\bm r_\theta = (x_\theta,y_\theta,z_\theta)$, and $\dot {\bm r_\theta}=(\dot x_\theta, \dot y_\theta, \dot z_\theta)$. Finally, using $\|\dot{\bm r}_\theta\|^2 = \|\dot{ \bm r}^\perp_\theta\|^2 + \|\dot{ \bm r}^=_\theta \|^2$ and $\|\dot{\bm r}_\theta \cross \bm r_\theta\|^2= \|\dot{ \bm r}^\perp_\theta\|^2 \|{ \bm r}_\theta\|^2$, we obtain 
\begin{equation}
{\rm QFI}(\rho_\theta) =\|\dot{ \bm r}^\perp_\theta\|^2 + \frac{ \|\dot{ \bm r}^=_\theta \|^2}{1-\|\bm r_\theta\|^2}
\end{equation}

\section{The reduced evolution in the $\alpha \ll 1$ limit}
\label{app: vacuum limit}
In this limit, we can expand the coherent state as
\begin{equation}
\ket{\alpha} = \left(1-\frac{\alpha^2}{2}\right)\ket{0} + \alpha \ket{1} +\frac{\alpha^2}{\sqrt{2}}\ket{2} + O(\alpha^3).
\end{equation}
With this expansion, the map can be decomposed as $G^{\alpha} = G^{(0)} + \alpha G^{(1)} +\alpha^2 G^{(2)}  + O(\alpha^3)$ with 
\begin{align}
  G^{(0)}  &=\left(
\begin{array}{cccc}
 1 & \cos (\tau ) & 0 & 0 \\
 \cos (\tau ) & \cos ^2(\tau ) & 0 & 0 \\
 0 & 0 & \sin ^2(\tau ) & 0 \\
 0 & 0 & 0 & 0 \\
\end{array}
\right),\\
G^{(1)}& = \left(
\begin{array}{cccc}
 0 & 0 & -\frac{1}{2} \sin (2 \tau ) & \sin (\tau ) \\
 0 & 0 & - \sin (\tau ) \cos \left(\sqrt{2} \tau \right) & \frac{1}{2} \sin (2 \tau ) \\
 -\frac{1}{2} \sin (2 \tau ) & - \sin (\tau ) \cos \left(\sqrt{2} \tau \right) & 0 & 0 \\
 \sin (\tau ) & \frac{1}{2} \sin (2 \tau ) & 0 & 0 \\
\end{array}
\right),\\
G^{(2)} & = \left(
\begin{array}{cccc}
 \cos ^2(\tau )-1 & \cos (\tau ) \left(\cos \left(\sqrt{2} \tau \right)-1\right) & 0 & 0 \\
 \cos (\tau ) \left(\cos \left(\sqrt{2} \tau \right)-1\right) & \cos ^2\left(\sqrt{2} \tau \right)-\cos ^2(\tau ) & 0 & 0 \\
 0 & 0 & \sin ^2\left(\sqrt{2} \tau \right)-\sin ^2(\tau ) & -\frac{\sin (\tau ) \sin \left(\sqrt{2} \tau \right)}{\sqrt{2}} \\
 0 & 0 & -\frac{\sin (\tau ) \sin \left(\sqrt{2} \tau \right)}{\sqrt{2}} & \sin ^2(\tau ) \\
\end{array}
\right),
\end{align}
and we recall that $\cE_{\tau|\alpha}[\bullet] =\sum_{i,j=0}^3 G^\alpha_{ij} \, L_i \bullet L_j^\dag, $ with $\{L_0,L_1,L_2,L_3\} =\{ \ketbra{0},\ketbra{1},\sigma_-,\sigma_+\}$. In particular, at $\alpha=0$ by diagonalizing the matrix $G^{(0)}$ we find the amplitude-damping (spontaneous-decay) channels 
\begin{align}
    \cE_{\tau|0}[\bullet] &= K_0 \bullet K_0^\dag + K_1 \bullet K_1^\dag \qquad \text{with}\\
    K_0 &= \sin(\tau) \sigma_-
    \qquad K_1 = \ketbra{0} + \cos(\tau)\ketbra{1},
\end{align}
where we recognize the vacuum Rabi oscillations for the initial state $\ket{e}$.

In general though, the state after applying the map introduced before on a real initial state $\rho_0 = \frac{1}{2}\left(\id + x_0 \sigma_x + z_0 \sigma_z\right)$ reads $\rho_{\tau|\alpha}=\cE_{\tau|\alpha}[\rho_0] =\frac{1}{2}\left(\id + x_\alpha\sigma_x + z_\alpha\sigma_z\right)$
with 
\begin{align}
    x_\alpha&=x_0 \cos (\tau ) \left(\alpha ^2 \cos \left(\sqrt{2} \tau \right)-\alpha ^2+1\right)+\frac{1}{2} \alpha  \sin (\tau ) \left(-\sqrt{2} \alpha  x_0 \sin \left(\sqrt{2} \tau \right)+2 (z_0-1) \cos \left(\sqrt{2} \tau \right)+2 z_0+2\right), \\
    z_\alpha&=\frac{1}{2} \left(-2 \alpha  x_0 \sin (2 \tau )+\alpha ^2 (z_0-1) \cos \left(2 \sqrt{2} \tau \right)+\cos (2 \tau ) \left(2 \alpha ^2+z_0-1\right)-\left(\alpha ^2-1\right) (z_0+1)\right).
\end{align}
From this, we can compute the QFI for any real initial state of the atom. However, to avoid cluttering, we will only give its expression for the atom initialized in the ground or in the excited state, i.e., $x_0=0$, and $z_0 = \pm 1$, 
\begin{align}
    {\rm QFI}[\rho^{(g)}_{\tau|\alpha}]&=4 \sin ^2(\tau ),\\
    {\rm QFI}[\rho^{(e)}_{\tau|\alpha}]&=\scalebox{1.1}{$\frac{\left(2 \alpha  \cos (2 \tau )-2 \alpha  \cos \left(2 \sqrt{2} \tau \right)\right)^2+4 \sin ^2(\tau ) \cos ^2\left(\sqrt{2} \tau \right)-4 \sin ^2(\tau ) \cos ^2\left(\sqrt{2} \tau \right) \left(\left(\alpha ^2+1\right) \cos (2 \tau )-\alpha ^2 \cos \left(2 \sqrt{2} \tau \right)\right)^2}{1-\left(\left(\alpha ^2-1\right) \cos (2 \tau )-\alpha ^2 \cos \left(2 \sqrt{2} \tau \right)\right)^2-4 \alpha ^2 \sin ^2(\tau ) \cos ^2\left(\sqrt{2} \tau \right)}$}.
\end{align}

\section{The reduced atomic dynamics and QFI computation}
\label{app: cutoff comp}

Before discussing the case of coherent field, let us say a few words about the reduced atomic state given by the channel $\cE$ in Eq.~\eqref{eq: channel in Gramm} in general. For any real initial state $\mathfrak{F}$ of the field mode, the elements of $G_{ij}$ are real, and the matrix is symmetric $G_{ij}=G_{ji}$. It is thus described by ten real parameters. By construction, the channel $\cE$ is trace preserving, hence it satisfies the identities
\begin{align}
\tr \cE(\id) =2 \qquad \tr \cE(\sigma_i) =0,
\end{align}
which can be used to derive the following constraints on the elements of $G_{ij}$: 
\be
G_{13}= -G_{02},\qquad G_{22}= 1-G_{11},\qquad \text{and} \qquad G_{33} =1-G_{00}.
\ee 
The atomic evolution is thus fully specified by seven real parameters $G_{00}, G_{11}, G_{03},G_{12},G_{01}, G_{02}, G_{23}$.

In particular, it follows that for any real initial state of the atom  $\rho_{0} =\frac{1}{2}\left(\id + x_0 \sigma_x + z_0 \sigma_z \right)$,  the final state is also real $\rho_{\tau|\alpha} =\frac{1}{2}\left(\id + x_{\tau|\alpha} \sigma_x + z_{\tau|\alpha} \sigma_z \right)$ and given by 
\begin{align} \label{eq: gen x}
x_{\tau|\alpha} &=  \left(G_{03}+G_{12}\right) +  x_0 \left(G_{01}+G_{23}\right) +  z_0 \left(G_{03}-G_{12}\right),\\
\label{eq: gen z}
z_{\tau|\alpha} &=  \frac{1}{2} \left(G_{00}-G_{11}+G_{22}-G_{33}\right) +  x_0 \left(G_{02}-G_{13}\right) + \frac{1}{2} z_0 \left(G_{00}+G_{11}-G_{22}-G_{33}\right)
\\
&= \left( G_{00}-G_{11} \right) +  x_0 \,2 \, G_{02} + z_0 \left(G_{00}+G_{11}-1 \right).
\end{align}

 When the initial state is ground or excited ($x_0=0, z_0=\pm1 $) we find that only the four real coefficients $G_{00}, G_{11},G_{03},G_{12}$ contribute to the final state
\begin{align}\label{eq app: x,z ge}
x_{\tau|\alpha}^{(g)} &=  2 \, G_{03}, &\quad z_{\tau|\alpha}^{(g)} &=   2 \,G_{00}-1,
\\ \nonumber
x_{\tau|\alpha}^{(e)} &=  2 \, G_{12}, &\qquad
z_{\tau|\alpha}^{(e)} &=   1 -2\, G_{11}.
\end{align}

We now focus on the reduced dynamic of the atom for coherent radiation initialized in the coherent state $\ket{\alpha}$, which has Poissonian photon number distribution $\ket{\alpha} = \sum_{n=0}^\infty \sqrt{{\rm P}(n|\alpha)}\ket{n}$ with ${\rm P}(n|\alpha)=\frac{\alpha^{2n}}{n!}e^{-\alpha^2}$. Recall that the Gram matrix is computed with the overlaps of the wave vectors
\begin{align}\label{eq app: state gens}
\ket{\alpha_0} &=\cos(\tau \sqrt{\hat n })\ket{\alpha}, &\qquad  \ket{\alpha_2} &=- \frac{\sin(\tau \sqrt{\hat n })}{\sqrt{\hat n}} a^\dag\ket{\alpha} = - \sin(\tau \sqrt{\hat n }) \frac{\sqrt{\hat n}}{\alpha}\ket{\alpha},
\\
\nonumber
\ket{\alpha_1} &=\cos(\tau \sqrt{\hat n+1 })\ket{\alpha}, &\quad 
\ket{\alpha_3} &= \frac{\sin(\tau \sqrt{\hat n+1})}{\sqrt{\hat n+1}} a \ket{\alpha} = \alpha \frac{\sin(\tau \sqrt{\hat n+1})}{\sqrt{\hat n+1}} \ket{\alpha},
\end{align}
where we used ${\rm P}(n|\alpha) = \frac{\alpha^2}{n} {\rm P}(n-1|\alpha)$ and
\begin{align}
a^\dag \ket{\alpha} &=a^\dag \sum_{n=0}^\infty \sqrt{{\rm P}(n|\alpha)}\ket{n} =  \sum_{n=0}^\infty \sqrt{{\rm P}(n|\alpha)}\sqrt{n+1}\ket{n+1} = \sum_{n=1}^\infty \sqrt{{\rm P}(n-1|\alpha)}\sqrt{n}\ket{n} = \sum_{n=0}^\infty \sqrt{{\rm P}(n|\alpha)}\frac{n}{\alpha}\ket{n}\\ &=
\frac{\hat n}{\alpha} \ket{\alpha}.
\end{align}
to rewrite the expression of $\ket{\alpha_2}$. The four states can be expressed as functions of the photon number operator applied on the coherent state $\ket{\alpha_i}=f_i(\hat n) \ket{\alpha}$, with 
\begin{align}\label{eq: state gens app}
f_0(\hat n) &=\cos(\tau \sqrt{\hat n }) &\quad \qquad  f_2(\hat n) &= - \sin(\tau \sqrt{\hat n }) \frac{\sqrt{\hat n}}{\alpha} 
\\
f_1(\hat n) &=\cos(\tau \sqrt{\hat n+1 })&\quad
f_3(\hat n)&= \sin(\tau \sqrt{\hat n+1}) \frac{\alpha}{\sqrt{\hat n+1}}. \nonumber
\end{align}
The entries of the Gram matrix can thus be expressed as follows
\be\label{eq: general G}
G_{ij}= \bra{\alpha} f_j(\hat n) f_i(\hat n) \ket{\alpha} = \sum_{n=0}^\infty {\rm P}(n|\alpha) f_i( n) f_j( n) = \mathds{E}[f_i(\hat n) f_j(\hat n)],
\ee
and computed as expected values of a function of a Poisson random variable $\hat n$, as given by the last expression.\\

Similarly, one can express the derivative of the Gram matrix $\dot G_{ij}:= \frac{\dd}{\dd \alpha} G_{ij}$ with respect to the parameter $\alpha$, which by linearity gives the derivative of the reduced state of the atom
\be
\dot \rho_{\tau|\alpha} = \sum_{i,j=0}^3 \dot G_{ij} \, L_i\,  \rho_{0} \, L_j^\dag,
\ee
required to compute the QFI. Using $\dot {\rm P}(n|\alpha) = 2\frac{n-\alpha^2}{\alpha} {\rm P}(n|\alpha)$ and 
\begin{align}
\dot f_0(\hat n) &=0 &\qquad  \dot f_2( n) &= \sin(\tau \sqrt{n }) \frac{\sqrt{n}}{\alpha^2} = -\frac{f_2( n)}{\alpha} 
\\
\dot f_1(n) &= 0 &\qquad 
\dot f_3(n)&= \frac{\sin(\tau \sqrt{ n+1})}{\sqrt{ n+1}} = \frac{f_3(n)}{\alpha},
\end{align} 
we obtain
\begin{align}\label{eq: G dot}
\dot G_{ij}&= \sum_{n=0}^\infty {\rm P}(n|\alpha) \left(\frac{2n-2\alpha^2}{\alpha} f_j( n) f_i( n)+\dot f_j( n) f_i( n)+ f_j( n) \dot f_i( n)\right) \\
&= \mathds{E}\left[\frac{2\hat n-2\alpha^2}{\alpha} f_j( \hat n) f_i( \hat n)+\dot f_j( \hat n) f_i(\hat  n)+ f_j( \hat n) \dot f_i( \hat n)\right]. \nonumber
\end{align}

Combining the Eqs.~(\ref{eq app: x,z ge}, \ref{eq: general G}, \ref{eq: G dot}) we find that for the atom initialized in the ground or excited states, the final state reads $\rho_{\tau|\alpha}^{(g/e)} =\frac{1}{2}\left(\id + x_\alpha^{(g/e)}\sigma_x + z_\alpha^{(g/e)}\sigma_z \right)$ with 
\begin{align} \label{eq: xg xdot g}
    x_\alpha^{(g)} &= 2\,\mathds{E}\left[ \frac{ \alpha}{\sqrt{\hat n +1}} \cos(\tau\sqrt{\hat n}) \sin(\tau\sqrt{\hat n+1}) \,  \right] &\quad 
    z_\alpha^{(g)} &= 2\,\mathds{E}\left[\cos^2(\tau\sqrt{\hat n})\right]-1 \\
\dot x_\alpha^{(g)} &=
 \, 2\, \mathds{E}\left[  \frac{ (2\hat n-2\alpha^2+1) }{\sqrt{\hat n+1}}\,\cos(\tau\sqrt{\hat n}) \sin(\tau\sqrt{\hat n+1}) \right]
 &\quad
\dot z_\alpha^{(g)} &= 2\,\mathds{E}\left[  \frac{2\hat n-2\alpha^2}{\alpha} \cos^2(\tau\sqrt{\hat n})\right] 
\end{align} 
for the ground state, and

\begin{align}
x_\alpha^{(e)} 
&= -2\,\mathds{E}\left[\frac{\sqrt{\hat n}}{\alpha} \cos(\tau\sqrt{\hat n+1}) \sin(\tau \sqrt{\hat n })\right] &\quad
z_\alpha^{(e)} &= 1-2\,\mathds{E}\left[ \cos^2 (\tau \sqrt{\hat n+1}) \right]\\
 \dot x_\alpha^{(e)} &= -2\, \mathds{E}\left[\frac{\sqrt{\hat n} (2\hat n-2\alpha^2-1) }{\alpha^2} \cos(\tau\sqrt{\hat n+1}) \sin(\tau \sqrt{\hat n })\right] &\quad
  \dot z_\alpha^{(e)} &= -2\, \mathds{E}\left[\frac{2\hat n- 2\alpha^2}{\alpha} \cos^2 (\tau \sqrt{\hat n+1}) \right]
\end{align}
for the excited. \\

\section{The atomic dynamics in the large $\alpha$ limit }
\label{app: large alpha limit}

Our goal is to compute the expressions of the form (Eq.~\ref{eq: general G})
\begin{align}    
G_{ij} &= \sum_{n=0}^\infty {\rm P}(n|\alpha) f_i( n) f_j( n) = \mathds{E}[f_i( \hat n) f_j( \hat n) ] 
\end{align}
for the seven values $G_{00}, G_{11}, G_{03}, G_{12}, G_{01}, G_{02}, G_{23}$ determining the reduced dynamics of the atom, where the functions $f_i$ are given in Eq.~\eqref{eq: state gens app} and $\hat n$ is a Poisson random variable with mean $\alpha^2$. Using the complex exponential representation of sine and cosine, all the products 
\be
f_i(n) f_j(n) = \Gamma_{ij}(n) \sum_{k,\pm} c_{i,j}(k,\pm) \,  e^{\pm \ii S_k(n)}
\ee 
can be expressed as linear combinations of some phases term $e^{ \pm \ii S_k(n)}$ with 
\be
\big(S_0(n),S_1(n), S_2(n), S_3(n)\big) = \big(2 \tau\sqrt{n},2 \tau \sqrt{n+1},\tau(\sqrt{n}+\sqrt{n+1}),\tau (\sqrt{n}-\sqrt{n+1}) \big)
\ee
times a factor $\Gamma_{ij}(n) \in \{1, \frac{\sqrt n}{\alpha},\frac{\alpha}{\sqrt {n+1}},\frac{\sqrt n}{\sqrt{n+1}}\}$,  and suitable weights $c_{i,j}(k,\pm)$. Note already, that by expanding in $\hat \delta = \frac{\hat n-\alpha^2}{\alpha}=O(1)$ we obtain $\Gamma_{ij}(n) = 1 + O(\alpha^{-1})$ in all the four cases. We are thus interested in computing the expressions of the form  
\begin{align}\label{eq: sum to compute}
  \sum_{n=0}^\infty P(n|\alpha) \Gamma_{ij}( n)e^{\ii S_k (n)} &=  \sum_{n=0}^\infty P(n|\alpha) e^{\ii S_k (n)} + O(\alpha^{-1}).
\end{align}

We will now separately discuss the limits where the phases $S_0(n), S_1(n), S_2(n)$ and $S_3(n)$ are approximately linear or not.

\subsection{The linear phase regime }

When the interaction time $\tau$ is relatively slow, the phases are well approximated by linear functions on the support of the Poissonian distributions. More precisely expanding in $\hat \delta = \frac{\hat n-\alpha^2}{\alpha}=O(1)$ one finds
\begin{align}
    S_0(n),S_1(n), S_2(n) &= 2 \alpha  \tau + \tau \frac{n-\alpha^2}{\alpha} + \tau O(\alpha^{-1}) \\
    S_3(n)&= -\frac{\tau }{2 \alpha } +\frac{  \tau }{4 \alpha ^2}\frac{n-\alpha^2}{\alpha} + \tau O(\alpha^{-3}).
\end{align}
Using the moment generating function of the Poissonian, it is then immediate to obtain 
\begin{align}
\label{eq: IK0}
\mathcal{I}_{k,0} &:= \sum_{n=0}^\infty P(n|\alpha) e^{\ii S_{k}(n)} = \exp(-\frac{\tau^2}{2}+\ii \, 2 \alpha\,  \tau \, ) \qquad \text{for } \, k=1,2,3.\\
\label{eq: I30}
\mathcal{I}_{3,0} &:=\sum_{n=0}^\infty P(n|\alpha) e^{\ii S_3(n)} = \exp(-\frac{\tau^2}{32 \alpha^4}-\ii \, \frac{\tau}{2\alpha}),
\end{align}
up to the order $O(\nicefrac{\tau}{\alpha})$ and $O(\nicefrac{\tau}{\alpha^3})$ respectively.

\subsection{The rapid phases regime}

In the limit where the linear phase approximation fails, to compute such sums in the right-hand side of Eq.~\eqref{eq: sum to compute}, we follow the approach presented in~\cite{fleischhauer1993revivals}. For simplicity we now focus on the phases for $k=0,1,2$ and come back to the case $k=3$ at the end of the section. First, we use the Poissonian summation~\cite{courant89} to express the discrete sum over $n$ with an integral by introducing the summation over an additional index
\be\label{eq: integrals}
\sum_{n=0}^\infty P(n|\alpha) e^{\ii S_k (n)} = \frac{1}{2}  P(0|\alpha) e^{\ii S_k(0)} + \sum_{\nu =-\infty}^\infty \int_0^\infty P(n|\alpha)  e^{\ii S_k(n) - \ii 2 \pi  n \nu}\dd n,
\ee
where the first term $P(0|\alpha)\approx 0$ can be already neglected in the limit of interest. Note that in the limit of interest the envelope $P(n|\alpha)$ is a function varying slowly, on the scale $n-n' =O(\alpha)$. In comparison, the phases $\varphi_{k,\nu}(n):= S_k(n) - 2 \pi  n \nu\approx 2\tau \sqrt{n}-2\pi n \nu $ are varying much faster. This is immediate to see for $\nu\neq 0$. For the case $\nu=0$, expanding in  $\hat \delta = \frac{\hat n-\alpha^2}{\alpha}=O(1)$, one notices that 
\be
\tau \sqrt{n} = \alpha \tau + \frac{\tau (n -\alpha^2)}{\alpha} + \tau O(\alpha^{-1})
\ee
also varies fast compared to the envelope in the limit of interest $\nicefrac{\tau}{\alpha} \geq O(1)$. Therefore the integrals in Eq.~\eqref{eq: integrals}
can be approximated with help of the stationary phase approximation, 
\begin{align}\label{eq: I3nu def}
\int_0^\infty P(n|\alpha)  e^{\ii \, \varphi_{k,\nu}(n)}\dd n \approx 
\mathcal{I}_{k,\nu}:= P(n_{k,\nu}|\alpha)  \exp(\ii \varphi_{k,\nu}(n_{k,\nu}) +\ii \frac{\pi}{4}{\rm sign}(\varphi''_{k,\nu}(n_{k,\nu}))) \sqrt{\frac{2\pi}{|\varphi''_{k,\nu}(n_{k,\nu})|}},
\end{align}
where $n_{k,\nu}\in[0,\infty]$ is the unique solution of 
\be
\varphi'_{k,\nu}(n_{k,\nu})=0 \quad \Longleftrightarrow \quad S'_k(n_{k,\nu}) = 2\pi \nu,
\ee
and is only a function of $\tau$ and $\nu$.  In all of the cases, we find that $S'_k(n)>0$ (on the support of the Poissonian), hence only the cases $\nu\geq 1$ give rise to a stationary phase $n_{k,\nu}$ inside the integration domain, and we only consider these values. Furthermore, we find that $
\varphi''_{k,\nu}(n) =S''_{k}(n) \leq 0$, giving
\be\label{eq: D17}
\mathcal{I}_{k,\nu}:= P(n_{k,\nu}|\alpha) e^{-\ii \nicefrac{\pi}{4}}\exp\big(\ii \, S_{k}(n_{k,\nu})- \ii 2\pi \nu \, n_{k,\nu} \big) \sqrt{\frac{2\pi}{|S''_{k}(n_{k,\nu})|}}.
\ee

For $k=0,1$ the the  stationary phase point $ n_{0,\nu} = \frac{\tau^2}{4\pi^2\nu^2 }$ and   $n_{1,\nu} = \frac{\tau^2}{4\pi^2 \nu^2 }-1$ are immediate to compute. For the remaining case $k=2$, the stationary phase point $n_{2,\nu}$ is the solution of $\frac{1}{\sqrt{n_{2,\nu}}}+\frac{1}{\sqrt{n_{2,\nu}+1}} =\frac{4 \pi \nu}{\tau}$. As $\frac{2}{\sqrt {n+1}}< \frac{1}{\sqrt{n}}+\frac{1}{\sqrt{n+1}} < \frac{2}{\sqrt n}$,  it follows that $n_{1,\nu}=n_{0,\nu}-1 < n_{2,\nu}< n_{0,\nu}$. Writing $n_{2,\nu} = n_{0,\nu}- \Delta =\frac{\tau^2}{4\pi^2\nu^2 } -\Delta$ with $\Delta =O(1)$, we find
\be\nonumber
0 =\frac{1}{\sqrt{n_{2,\nu}}}+\frac{1}{\sqrt{n_{2,\nu}+1}} -\frac{4 \pi \nu}{\tau} = \frac{1}{2} (2 \Delta -1) \left(\frac{1}{n_{0,\nu}}\right)^{3/2} + O(n_{0,\nu}^{-\nicefrac{5}{2}}),
\ee
showing that $\Delta=\frac{1}{2}$ and $n_{2,\nu} = \frac{\tau^2}{4\pi^2 \nu^2 }-\nicefrac{1}{2}$ in the limit of interest.\\

Plugging these values in Eq.~\eqref{eq: D17} we immeadiately find
\begin{align}
\mathcal{I}_{k,\nu} &= P\left(n_{k,\nu}\middle|\alpha\right)  e^{-\ii \nicefrac{\pi}{4}} \exp(\ii \frac{\tau^2}{2\pi \nu}+\ii \pi \nu c_k ) \frac{\tau}{\sqrt{2} \pi \nu^{\nicefrac{3}{2}}},
\end{align}
 with $c_0=0, c_1=2$ and $c_2=1$. Next, considering  the normal approximation of the square root of a Poissonian random variable~\cite{johnson2005univariate,fleischhauer1993revivals}
\be\label{eq: normal approx}
P(n_{k,\nu}|\alpha) \approx \frac{1}{\sqrt{2\pi} \alpha }e^{-2(\sqrt{n_{k,\nu}}-\alpha)^2 } \approx\frac{1}{\sqrt{2\pi} \alpha }\exp(-2\left(\frac{\tau}{2\pi\nu }-O(\nicefrac{\nu}{\tau})- \alpha \right)^2 ),
\ee
we see that our expressions are only non-vanishing when  $\frac{\tau}{2\pi\nu }-\alpha = O(1)$ (which implies $\frac{\nu}{\tau}\to 0$). Writing  $\tau = 2\pi \nu \alpha + \delta\tau$ with $\delta\tau = O(1)$ we obtain 
\be
P(n_{k,\nu}|\alpha) \approx \frac{1}{\sqrt{2\pi} \alpha}
\exp(-\frac{\delta \tau^2}{2\pi^2\nu^2}) = \frac{1}{\sqrt{2\pi} \alpha}
\exp(-\frac{(\tau -  2\pi \nu \alpha )^2}{2\pi^2\nu^2}).
\ee
This leads to simple asymptotic expressions
\begin{align}\label{eq:I0all}
\mathcal{I}_{0,\nu} =\mathcal{I}_{1,\nu}=(-1)^\nu \mathcal{I}_{2,\nu}  &= \frac{1}{\sqrt{\pi \nu}}
\exp(-\frac{(\tau -  2\pi \nu \alpha )^2}{2\pi^2\nu^2}) \exp(\ii  (\frac{\tau^2}{2\pi \nu} -\frac{\pi}{4}))
\end{align}
where we used $\exp(\ii 2\pi \nu )=1$ and $\exp(\ii \pi \nu )=(-1)^\nu$.

 \subsubsection{Stationary phase approximation for $S_3(n)$}
 
 To find the stationary phase points $n_{3,\nu}$  for the phase $S_3(n)$ one has to solve the following equation $\frac{1}{\sqrt{n_{3,\nu}}}-\frac{1}{\sqrt{n_{3,\nu}+1}} =\frac{4 \pi \nu}{\tau}$. Knowing that $n_{3,\nu}$ is large and expanding the left-hand side, we find that the solution is given by
 \be
 n_{3,\nu} =\frac{\tau^{2/3}}{4 (\pi \nu)^{2/3}} -\frac{1}{2} + o(1).
 \ee
Now we can use the stationary phase approximation to compute $\mathcal{I}_{3,\nu}$ via Eq.~\eqref{eq: I3nu def}. Similar to the other cases, we find  
\begin{align}
\varphi_{3}(n_{3,\nu})&= S_{3}(n_{3,\nu}) -2\pi \nu n_{3,\nu}=
-\frac{3}{2} \left(\pi \nu \right)^{1/3} \tau ^{2/3}  + \pi  \nu + O(\tau^{-\nicefrac{1}{3}}), \\
 S_3''(n_{3,\nu})&= -12 \frac{(\pi\nu)^{5/3}}{\tau ^{2/3}}+ O(\tau^{-\nicefrac{4}{3}}), \\
 \frac{1}{\sqrt{2 \pi}\alpha} \sqrt{\frac{2 \pi}{|S''_{k}(n_{3,\nu})|}} &=  \frac{1}{\sqrt{3 \pi \nu } } +O(1),
\end{align}
where we used that $\alpha = \sqrt{\frac{\tau^{2/3}}{4 (\pi \nu)^{2/3}}+O(1)}$ guaranteed by the exponential decay of $P(n_{3,\nu}|\alpha)$. Finally, plugging everything in Eq.~\eqref{eq: normal approx} and using the normal approximation~\eqref{eq: normal approx} we obtain the final expression 
\begin{align}
\mathcal{I}_{3,\nu} &= \frac{1}{\sqrt{3 \pi \nu}} \exp(-2 \left(\frac{\tau^{1/3}}{2 (\pi \nu)^{1/3}} -\alpha \right)^2)\exp(-\ii \nicefrac{\pi }{4})(-1)^\nu\exp( -\ii \frac{3}{2} \left(\pi \nu \right)^{1/3} \tau ^{2/3}) \\
&= \frac{1}{\sqrt{3 \pi \nu}} \exp(-\frac{\left(\tau^{1/3}-2  (\pi  \nu )^{1/3} \alpha \right)^2}{2 (\pi  \nu )^{2/3}})(-1)^\nu\exp(-\ii \frac{3}{2} \left(\pi \nu \right)^{1/3} \tau ^{2/3} - \ii \frac{\pi}{4}),
\end{align}
exhibiting revivals at times $\tau \approx 8 \pi \nu  \alpha ^3$.

\subsection{The full reduced dynamics of the atom}

Summarizing the above calculations, we find that in the limit of large $\alpha$ with $\nicefrac{\tau}{\alpha^3}\to 0$ and up to $O(\alpha^{-1})$ the expected values in Eq.~\eqref{eq: sum to compute} are given by 
\be
 \sum_{n=0}^\infty P(n|\alpha) \Gamma_{ij}( n)e^{\ii S_k (n)} \approx
\begin{cases}
\sum_{\nu=0}^\infty \mathcal{I}_{0,\nu} & k=0,1    \\ 
\sum_{\nu=0}^\infty \mathcal{I}_{0,\nu} (-1)^\nu& k=2
\\
\sum_{\nu=0}^\infty \mathcal{I}_{3,\nu}& k=3
\end{cases},
\ee
with all the terms given in Eqs.~(\ref{eq:I0all}, \ref{eq: I30}, \ref{eq: IK0}). This allows us to express the matrix $G$, representing the channels, as 
\begin{align}
G &=\frac{1}{2}\id + \sum_{\nu=0}^\infty (\cG_\nu^{(0)} +\cG_\nu^{(3)}) \qquad \text{with}\\
\cG_{\nu}^{(0)} &=\frac{1}{2} 
\left( \begin{array}{cccc}
\cI_{0,\nu}^{\rm Re} & (-1)^\nu\, \cI_{0,\nu}^{\rm Re} & -  \cI_{0,\nu}^{\rm Im} & (-1)^\nu \,\cI_{0,\nu}^{\rm Im}\\
(-1)^\nu\, \cI_{0,\nu}^{\rm Re}& \cI_{0,\nu}^{\rm Re} & -(-1)^\nu \,\cI_{0,\nu}^{\rm Im}  & \cI_{0,\nu}^{\rm Im}\\
- \cI_{0,\nu}^{\rm Im}&- (-1)^\nu\,\cI_{0,\nu}^{\rm Im}&-\cI_{0,\nu}^{\rm Re} & (-1)^\nu \,\cI_{0,\nu}^{\rm Re}\\
 (-1)^\nu \,\cI_{0,\nu}^{\rm Im}& \cI_{0,\nu}^{\rm Im}& (-1)^\nu \, \cI_{0,\nu}^{\rm Re}& -\cI_{0,\nu}^{\rm Re}
\end{array}\right) \qquad
\cG_{\nu}^{(3)} =\frac{1}{2}
\left( \begin{array}{cccc}
& \cI_{3,\nu}^{\rm Re} &  &- \cI_{3,\nu}^{\rm Im} \\
\cI_{3,\nu}^{\rm Re}&  & -\cI_{3,\nu}^{\rm Im} & \\
& -\cI_{3,\nu}^{\rm Im} & &-\cI_{3,\nu}^{\rm Re}\\ 
- \cI_{3,\nu}^{\rm Im} & & -\cI_{3,\nu}^{\rm Re} & 
\end{array}\right)
\\
\cI_{0,\nu} &= 
\begin{cases}\exp(-\frac{\tau^2}{2}+\ii \, 2 \alpha\,  \tau \, )& \nu =0 \\
\frac{1}{\sqrt{\pi \nu}}
\exp(-\frac{(\tau -  2\pi \nu \alpha )^2}{2\pi^2\nu^2}) \exp(\ii  (\frac{\tau^2}{2\pi \nu} -\frac{\pi}{4})) & \nu \geq 1
\end{cases}\\
\cI_{3,\nu}  &= \begin{cases}\exp(-\frac{\tau^2}{32 \alpha^4}-\ii \, \frac{\tau}{2\alpha}) & \nu =0 \\
 \frac{1}{\sqrt{3 \pi \nu}} \exp(-\frac{\left(\tau^{1/3}-2  (\pi  \nu )^{1/3} \alpha \right)^2}{2 (\pi  \nu )^{2/3}})(-1)^\nu\exp( -\ii \frac{3 (\pi \nu)^{\nicefrac{1}{3}}\tau^{\nicefrac{2}{3}}}{2} - \ii \frac{\pi}{4})  & \nu\geq 1
\end{cases}.
\end{align}
 All matrices $\cG_\nu^{(k)}$  correspond to the linear maps $\cE_\nu^{(k)}$, which are neither positive nor trace preserving. In contrast, the first matrix $\bar \cG  :=\frac{1}{2}\id$ corresponds to the fully depolarizing channel
\be
\bar \cE[\bm \sigma] =  \left(\begin{array}{c}
\id \\
0
\\
0
\\
 0
\end{array}\right) \quad \implies \quad 
\bar \cE[\rho] =  \frac{\id}{2} ,
\ee
which is the long time limit $\tau\gg \alpha^3$ of the atomic dynamics, where all the other terms vanish. For the other maps, we find
\begin{align}
\cE_\nu^{(0)}[\bm \sigma] = 
\left(\begin{array}{c}
0 \\
(-1)^\nu \, \cI_{0,\nu}^{\rm Re}\, \sigma_x - \cI_{0,\nu}^{\rm Im}\, \sigma_z
\\
0
\\
 (-1)^\nu \cI_{0,\nu}^{\rm Im}\, \sigma_x + \cI_{0,\nu}^{\rm Re}\, \sigma_z 
\end{array}\right), 
\qquad \qquad
\cE_\nu^{(3)}[\bm \sigma] = 
\left(\begin{array}{c}
- \cI_{3,\nu}^{\rm Im}\, \sigma_x \\
0
\\
\cI_{3,\nu}^{\rm Re} \, \sigma_y
\\
0
\end{array}\right),
\end{align}
where $\bm \sigma = (\id, \sigma_x,\sigma_y,\sigma_z)$.
With the help of these expressions, we now discuss the exact form of the channel $\cE_{\tau|\alpha}$ in different time limits.\\

{\bf Short times $\tau <O(\alpha)$.} First, consider the short time limit $\tau <O(\alpha)$. Here, the only non-vanishing terms are $\cI_{3,0}= 1$ and $\cI_{0,0}= \exp(-\frac{\tau^2}{2}+\ii \, 2 \alpha\,  \tau \, )$, implying
\be\label{eq app: channel short}
\tau\leq O(\alpha): \qquad \cE_{\tau|\alpha}[\bm \sigma] =  
\left(\begin{array}{c}
\id \\
e^{-\nicefrac{\tau^2}{2}}\Big(\cos(2 \alpha \,\tau)\, \sigma_x - \sin(2 \alpha \,\tau)\, \sigma_z\Big)
\\
\sigma_y
\\
e^{-\nicefrac{\tau^2}{2}}\Big(\sin(2 \alpha \,\tau)\, \sigma_x + \cos(2 \alpha \,\tau)\, \sigma_z\Big)
\end{array}\right).
\ee
\vspace{1 cm}

{\bf Longer times $1\ll \tau \ll \alpha^3$.} In this regime the term $\mathcal{I}_{0,0}$ and $\mathcal{I}_{3,\nu\geq 1}$ are negligible. Hence the atomic evolution is given by $G = \frac{1}{2} \id + \cG_0^{(3)} + \cG_\nu^{(0)}$ 
with 
\be\label{eq app: channel longer}
1\ll \tau\ll \alpha^3: \qquad \cE_{\tau|\alpha}[\bm \sigma] =  
\left(\begin{array}{c}
\id + e^{-\nicefrac{\tau^2}{32 \alpha^4}} \sin(\nicefrac{\tau}{2\alpha}) \, \sigma_x\\
\frac{1}{\sqrt{\pi \nu}}
\exp(-\frac{(\tau -  2\pi \nu \alpha )^2}{2\pi^2\nu^2}) \Big((-1)^\nu \cos(\frac{\tau^2}{2\pi \nu} -\frac{\pi}{4})\, \sigma_x - \sin(\frac{\tau^2}{2\pi \nu} -\frac{\pi}{4})\, \sigma_z \Big)
\\
e^{-\nicefrac{\tau^2}{32 \alpha^4}} \cos(\nicefrac{\tau}{2\alpha}) \,\sigma_y
\\
\frac{1}{\sqrt{\pi \nu}}
\exp(-\frac{(\tau -  2\pi \nu \alpha )^2}{2\pi^2\nu^2}) \Big((-1)^\nu \sin(\frac{\tau^2}{2\pi \nu} -\frac{\pi}{4}) \, \sigma_x + \cos(\frac{\tau^2}{2\pi \nu} -\frac{\pi}{4}) \, \sigma_z \Big)
\end{array}\right)
\ee
and $\nu$ can be taken to minimize $|\tau-2\pi \alpha\nu|$. Remarkably, $\cE_{\tau|\alpha}[\sigma_x]$ and $\cE_{\tau|\alpha}[\sigma_z]$  are only nonzero for $|\tau - 2\pi \nu \alpha|= O(1)$, where $\sin(\nicefrac{\tau}{2\alpha})\approx 0$ and $\cos(\nicefrac{\tau}{2\alpha})\approx (-1)^\nu$.  Hence, this time regime exhibits two very different behaviors: close and outside of revivals: \begin{align}
|\tau - 2\pi \nu \alpha|= O(1) \qquad &\implies \qquad \cE_{\tau|\alpha}[\id]=\id, \, \cE_{\tau|\alpha}[\sigma_y]= (-1)^\nu e^{-\nicefrac{\tau^2}{32 \alpha^4}}\sigma_y,\\
|\tau - 2\pi \nu \alpha|\gg 1 \quad &\implies \qquad \cE_{\tau|\alpha}[\sigma_x]=\cE_{\tau|\alpha}[\sigma_z]=0.
\end{align}

\vspace{1 cm}

{\bf Very long times $\tau \gg \alpha^2$:}  Finally at even longer times $\tau\gg \alpha^2$, the only non-vanishing contribution is $\cG_\nu^{(3)}$ with $\nu \geq 1$, showing revivals around specific times with 
\be\label{eq: channel very long}
|\tau^{1/3}-2 \alpha (\pi  \nu)^{1/3}|=O(1): \qquad \cE_{\tau|\alpha}[\bm \sigma] =  
\left(\begin{array}{c}
\id + \frac{1}{\sqrt{3 \pi \nu}}  e^{-F^2_\nu(\tau,\alpha)} (-1)^\nu\sin(\frac{3}{2} (\pi \nu \tau^2)^{1/3} + \frac{\pi}{4})  \, \sigma_x\\
0
\\
\frac{1}{\sqrt{3 \pi \nu}}  e^{-F^2_\nu(\tau,\alpha)} (-1)^\nu \cos(\frac{3}{2} (\pi \nu \tau^2)^{1/3} + \frac{\pi}{4}) \,\sigma_y
\\
0
\end{array}\right)
\ee
and $F_\nu(\tau,\alpha)=\frac{\tau^{1/3}-2 \alpha (\pi  \nu)^{1/3}}{\sqrt 2 (\pi  \nu)^{1/3}}$.

\subsection{The atomic state and its QFI}
\label{app: revival-dynamics}

To discuss the final state for an atom prepared in the ground or excited state and its QFI, we again separately consider the three different time regimes

\subsubsection{Quantum Fisher information at short times $\tau=O(1)$}

For $\tau\ll \alpha$ using Eq.~\eqref{eq app: channel short} one immediate obtains $x_{\tau|\alpha}^{(g/e)}= \pm e^{-\nicefrac{\tau^2}{2}} \sin(2\alpha \tau)$,  $z_{\tau|\alpha}^{(g/e)}= \pm e^{-\nicefrac{\tau^2}{2}} \cos(2\alpha \tau)$ (where the sign $\pm$ is determined by the initial condition), and 
\be
{\rm QFI}[\rho_{\tau|\alpha}^{(g/e)}] = 4 \tau^2 e^{-\tau^2}.
\ee
The QFI reaches its maximum at $\tau=1$ and then quickly decays to zero, together with the coherence and the population.

Let us also compute the Fisher information ${\rm FI}_z[\rho_{\tau|\alpha}^{(g/e)}]$for the population measurement of the atom. This is equivalent to computing the QFI of the atomic state, dephased in the $z$-basis, i.e., setting $x=\dot x =0$.  
\be
{\rm FI}_z[\rho_{\tau|\alpha}^{(g/e)}]= 4 \tau ^2 \frac{ \sin ^2(2 \alpha  \tau )}{e^{\tau ^2}-\cos ^2(2 \alpha  \tau )} \leq 4 \tau^2 e^{-\tau^2},
\ee
saturated for $ \sin ^2(2 \alpha  \tau )$. It is equal to the QFI of the state, for well-chosen times.

\subsubsection{Quantum Fisher information at long times $\tau=O(\alpha)$}

At longer times $1\ll \tau\ll \alpha^2$ by Eq.~\eqref{eq app: channel longer} the atomic state is found to be 
\begin{align} \label{eq: z get}
    z_{\tau|\alpha}^{(g/e)} & =  \pm \sum_{\nu= 1}^\infty \frac{1}{\sqrt{\pi \nu}}  e^{-R^2_\nu(\tau|\alpha)}\cos(\Phi_{\nu}) \\
     x_{\tau|\alpha}^{(g/e)}  &= e^{-\nicefrac{\tau^2}{32 \alpha^4}} \sin(\nicefrac{\tau}{2\alpha}) \pm \sum_{\nu = 1}^\infty \frac{1}{\sqrt{\pi \nu}} e^{-R^2_\nu(\tau|\alpha)} (-1)^\nu\sin(\Phi_{\nu}), \label{eq: x gen}
\end{align}
where we introduced $R_\nu(\tau|\alpha) := \frac{\tau -2 \pi  \alpha  \nu }{\sqrt{2} \pi \nu}$, $\Phi_\nu = \frac{\tau^2}{2\pi \nu}-\frac{\pi}{4}$ to shorten the equations. Deriving with respect to the parameter $\alpha$, one finds 
\begin{align}
    \dot z_{\tau|\alpha}^{(g/e)} & =  \pm \sum_{\nu= 1}^\infty \frac{2 \sqrt 2 \, R_\nu(\tau|\alpha)\, }{\sqrt{\pi \nu}}e^{-R^2_\nu(\tau|\alpha)}  \cos(\Phi_{\nu}) \\
     \dot x_{\tau|\alpha}^{(g/e)}  &= - \frac{\tau}{2 \alpha^2} e^{-\nicefrac{\tau^2}{32 \alpha^4}} \left(\cos(\frac{\tau}{2\alpha})-\frac{\tau}{4 \alpha^3}\sin(\frac{\tau}{2\alpha})\right) \pm \sum_{\nu= 1}^\infty \frac{2 \sqrt 2 \, R_\nu(\tau|\alpha)\, }{\sqrt{\pi \nu}}e^{-R^2_\nu(\tau|\alpha)}  (-1)^{\nu}\sin(\Phi_{\nu}),
     \label{eq: x dot gen}
\end{align}
where the term $\frac{\tau}{4 \alpha^3}\sin(\frac{\tau}{2\alpha})$ can be neglected. To compute the QFI, we discuss the times at revivals and outsider separately.\\

{\bf At revivals $|\tau - 2\pi \nu \alpha| = O(1)$}  First, let us consider the expressions of QFI close to the revival times $\tau \approx 2\pi \alpha \nu$, i.e., with $\delta\tau =\tau-2\pi \nu\alpha = O(1)$. For $\tau\ll \alpha^2$, the terms coming with the prefactor $e^{-\nicefrac{\tau^2}{32 \alpha^4}}$ can be neglected. The atomic dynamics is given by 
\begin{align}\nonumber
    z_{\tau|\alpha}^{(g/e)} & =  \pm  \frac{1}{\sqrt{\pi \nu}}  e^{-R^2_\nu(\tau|\alpha)}\cos(\Phi_{\nu}) & \quad \dot z_{\tau|\alpha}^{(g/e)} & =  \pm  \frac{2 \sqrt 2 \, R_\nu(\tau|\alpha)\, }{\sqrt{\pi \nu}}e^{-R^2_\nu(\tau|\alpha)}  \cos(\Phi_{\nu}) \\
     x_{\tau|\alpha}^{(g/e)}  &= \pm  \frac{1}{\sqrt{\pi \nu}} e^{-R^2_\nu(\tau|\alpha)} (-1)^\nu\sin(\Phi_{\nu}) &\quad 
     \dot x_{\tau|\alpha}^{(g/e)}  &= \pm \frac{2 \sqrt 2 \, R_\nu(\tau|\alpha)\, }{\sqrt{\pi \nu}}e^{-R^2_\nu(\tau|\alpha)}  (-1)^{\nu}\sin(\Phi_{\nu})
     \nonumber
\end{align}
leading to 
\be \label{eq: QFI at revivals}
{\rm QFI}[\rho^{(g/e)}_{\tau|\alpha}] 
=\frac{8 R_\nu^2(\tau|\alpha) }{\pi  \nu \,  e^{2 R_\nu^2(\tau|\alpha)}-1}.
\ee

\begin{wrapfigure}{h}{0.37\textwidth}
    \begin{center}
    \vspace{-0.4 cm}
    \includegraphics[width=0.36 \textwidth]{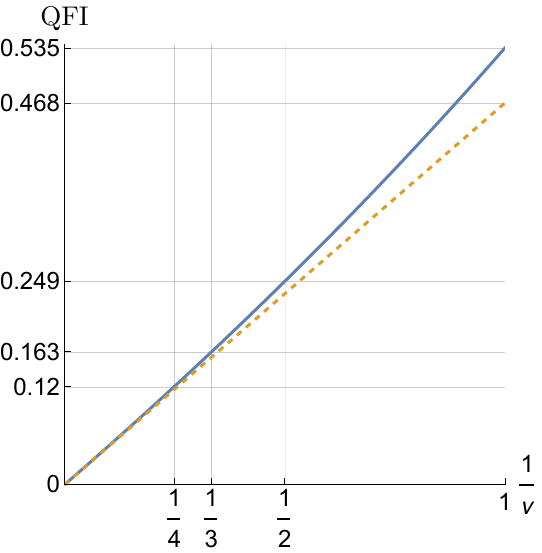}
    \end{center}
     \vspace{-0.2 cm}
    \caption{The function $-4 W_0\left(-\frac{1}{ e \pi  \nu }\right) $ (full line) and the linear lower bound $\frac{0.468}{ \nu }$ (dashed line).}
    \vspace{-0.5 cm}
    \label{fig:ProdLog}
\end{wrapfigure}

By fine-tuning the time $\tau$ this expression can be maximized with respect to $ R_\nu(\tau|\alpha)$. Doing so, we find the following optimal value  
\be
{\rm QFI}[\rho^{(g/e)}_{\tau|\alpha}] =  -4 W_0\left(-\frac{1}{ e \pi  \nu }\right)   \approx \frac{0.468}{ \nu },
\ee
where $W_0$ is the principal branch of the Lambert $W$ function. This linear approximation works well for large $\nu$, as illustrated in Fig.~\ref{fig:ProdLog}.  The bound is saturated at 
\be
\tau = 2 \pi  \alpha  \nu \pm \pi  \nu  \sqrt{1+W_0\left(-\frac{1}{e \pi  \nu }\right)} \approx 2 \pi  \alpha  \nu  \pm  \sqrt{1-\frac{0.12}{\nu }} \pi \nu \approx 2 \pi  \alpha  \nu \pm \pi \nu 
\ee
For $\nu=1$ or 2 the QFI can be as high as ${\rm QFI}[\rho^{(g/e)}_{\tau|\alpha}]\approx 0.54$ and 0.25.

For the Fisher information of the population measurement, we find 
\be
{\rm FI}_z[\rho_{\tau|\alpha}^{(g/e)}]= \frac{8 R_\nu^2(\tau|\alpha) \cos^2(\Phi_\nu) }{\pi  \nu \,  e^{2 R_\nu^2(\tau|\alpha)}-\cos^2(\Phi_\nu)},
\ee
which is equal to the full QFI in Eq.~\eqref{eq: QFI at revivals} at times with $\cos^2(\Phi_\nu)=1$. Note that $\Phi_\nu$ changes much faster than $R_\nu$, and can thus be optimized independently.\\

{\bf Outside of revivals $|\tau - 2\pi \nu \alpha|\gg 1$.}  For generic times $\tau-2\pi \nu\alpha > O(1)$ the contributions of the $\nu$ terms are vanishing, and the atomic state is given by $z_{\tau|\alpha}^{(g/e)}=0$ and 
\begin{align}
   x_{\tau|\alpha}^{(g/e)} &= e^{-\nicefrac{\tau^2}{32 \alpha^4}} \sin(\nicefrac{\tau}{2\alpha}) \qquad \dot x_{\tau|\alpha}^{(g/e)}  = - \frac{\tau}{2 \alpha^2} e^{-\nicefrac{\tau^2}{32 \alpha^4}} \cos(\frac{\tau}{2\alpha})
\end{align}
in the leading order of $\nicefrac{1}{\alpha}$. For $\nicefrac{\tau}{\alpha^3}\to 0$, the second term in $\dot x_{\tau|\alpha}^{(g/e)} $ can be neglected and  the quantum Fisher information is given by 
\be\label{eq: QFI of x}
{\rm QFI}[\rho^{(g/e)}_{\tau|\alpha}] = \frac{\tau^2}{4\, \alpha^4} \frac{\cos ^2\left(\frac{\tau }{2 \alpha }\right)}{\, e^{\frac{\tau ^2}{16 \alpha ^4}}-\sin ^2\left(\frac{\tau }{2 \alpha }\right)}.
\ee
For relatively short times $\tau \ll \alpha^2$ this expression is generically suppressed QFI $\approx 0$  by the prefactor.\\

Special care must be taken for times where the state is approximately pure, which occurs at $\tau \approx 2\pi \alpha (\nu - \nicefrac{1}{2})$ where $x^{(e/g)}_{\alpha|\tau} = \sin(\nicefrac{\tau}{2\alpha}) \approx 1$. Specifically, let us consider $\tau = \pi \alpha +\delta \tau$ with $\delta \tau = O(1)$, and focus on the atomic coherence for the ground initial state. With the help of Eqs.~\eqref{eq: xg xdot g}, we obtain 
\begin{align}
x_{\tau|\alpha}^{(g)} &=  \sum_{n=0}^\infty P(n|\alpha)\frac{ \alpha}{\sqrt{\hat n +1}} \sin(\tau(\sqrt{n+1}-\sqrt{\hat n}))  
\\
\dot x_{\tau|\alpha}^{(g)} &=  \sum_{n=0}^\infty P(n|\alpha)\frac{ (2\hat n-2\alpha^2+1) }{\sqrt{\hat n+1}} \sin(\tau(\sqrt{n+1}-\sqrt{\hat n}))  
\end{align}
by neglecting the fast phases. In this case, by taking the expansions to the next order, we have
\begin{align}
\sin(\tau(\sqrt{\hat n+1}-\sqrt{\hat n}))   &= 1- \frac{(\pi  \delta -2 \delta \tau )^2}{32 \alpha ^2} +O(\alpha^{-3})\\
\frac{\alpha}{\sqrt{\hat n+1}} &=1-\frac{\delta }{2 \alpha }+\frac{3 \delta ^2-4}{8 \alpha ^2}+  O(\alpha^{-3}) \\
\frac{ (2\hat n-2\alpha^2+1) }{\sqrt{\hat n+1}}  &= 2 \delta+ \frac{1-\delta ^2}{\alpha }+\frac{3 \left(\delta ^3-2 \delta \right)}{4 \alpha ^2}+ O(\alpha^{-3})
\end{align}
leading to 
\begin{align}\begin{cases}
x_{\tau|\alpha}^{(g)} =1-\frac{4 \delta \tau ^2+\pi ^2+4}{32 \alpha ^2} + O(\alpha^{-3}) \\
\dot x_{\tau|\alpha}^{(g)} = \frac{\pi  \delta \tau }{4 \alpha ^2}+ O(\alpha^{-3})
\end{cases} \implies \, {\rm QFI}[\rho^{(g)}_{\tau|\alpha}] = O(\alpha^{-2}).
\end{align}
The same conclusion  ${\rm QFI}[\rho^{(e)}_{\tau|\alpha}]=O(\alpha^{-2})$ can be repeated for the initially excited atom, showing that the QFI is indeed vanishing for $\alpha = O(\tau)$ outside of the revival times. 

Contrasting this asymptotic conclusion, we see that the numerical results of Fig.~\ref{fig: qfiground} show a nonzero ${\rm QFI}[\rho^{(e)}_{\tau|\alpha}]$ around $\tau \approx \pi \alpha$ for the excited initial condition. However, this effect disappears if we numerically compute the contribution to the QFI solely coming from the atomic coherence $\frac{(\dot x_{\tau|\alpha}^{(e)})^2}{1-(x_{\tau|\alpha}^{(g)})^2}$. This suggests that the nonzero quantum Fisher information it is due to a residual nonzero $\dot z_{\tau|\alpha}^{(e)}$ contribution to the numerator.
\\

\subsubsection{Quantum Fisher information at longer times $\tau=O(\alpha^2)$}

Here the expressions in Eqs.~(\ref{eq: z get}-\ref{eq: x dot gen}) are still valid, but for longer times $\nicefrac{\tau}{\alpha^2}=O(1)$, the terms with $ e^{-\nicefrac{\tau^2}{32 \alpha^4}}$ can not be ignored. In this regime, only the ``late'' revivals are relevant, as $\tau - 2\pi \alpha \nu =0$ implies $\nu = \frac{\tau}{2 \alpha}= O(\alpha)$. Moreover, the subsequent revivals stop being well separated in time
\be 
R_\nu(\tau=2\pi \alpha (\nu+1)|\alpha) = \frac{\tau-2\pi \alpha \nu}{\sqrt{2} \pi \nu}=\frac{2\pi \alpha (\nu+1)-2\pi \alpha \nu}{\sqrt{2} \pi \nu} = \sqrt 2 \, \nicefrac{\alpha}{\nu} = O(1).
\ee
In fact, there is a $O(1)$ number of integer values $\nu$ giving non-vanishing contributions at a fixed time $\tau$, we will denote this set with ${\rm N}_\tau$. Nevertheless, because of the $\nicefrac{1}{\sqrt{\nu}}$ prefactor in the expressions of the atomic state, the contributions of the revivals can be treated as a small perturbation
\begin{align}
    z_{\tau|\alpha}^{(g/e)} & =  0+\delta z, &\quad \delta z &:=\pm \sum_{\nu\in {\rm N}_\tau} \frac{1}{\sqrt{\pi \nu}}  e^{-R^2_\nu(\tau|\alpha)}\cos(\Phi_{\nu}), \\
     x_{\tau|\alpha}^{(g/e)}  &= e^{-\nicefrac{\tau^2}{32 \alpha^4}} \sin(\nicefrac{\tau}{2\alpha}) + \delta x, &\quad \delta x &:= \pm \sum_{\nu\in {\rm N}_\tau} \frac{1}{\sqrt{\pi \nu}} e^{-R^2_\nu(\tau|\alpha)} (-1)^\nu\sin(\Phi_{\nu}), \\
    \dot z_{\tau|\alpha}^{(g/e)} & = 0+  \delta \dot z, &\quad \delta \dot z &:=\pm \sum_{\nu\in {\rm N}_\tau}\frac{2 \sqrt 2 \, R_\nu(\tau|\alpha)\, }{\sqrt{\pi \nu}}e^{-R^2_\nu(\tau|\alpha)}  \cos(\Phi_{\nu}), \\
     \dot x_{\tau|\alpha}^{(g/e)}  &= - \frac{\tau}{2 \alpha^2} e^{-\nicefrac{\tau^2}{32 \alpha^4}} \cos(\frac{\tau}{2\alpha}) +\delta \dot x, &\quad \delta \dot x &:=\pm \sum_{\nu\in {\rm N}_\tau} \frac{2 \sqrt 2 \, R_\nu(\tau|\alpha)\, }{\sqrt{\pi \nu}}e^{-R^2_\nu(\tau|\alpha)}  (-1)^{\nu}\sin(\Phi_{\nu}),
\end{align}
where we neglected the second terms in  $\dot x_{\tau|\alpha}^{(g/e)} $  using $\nicefrac{\tau}{\alpha^3}\to 0$.
In the leading order, we recover
\be
{\rm QFI}[\rho^{(g/e)}_{\tau|\alpha}] = \frac{\tau^2}{4\, \alpha^4} \frac{\cos ^2\left(\frac{\tau }{2 \alpha }\right)}{\, e^{\frac{\tau ^2}{16 \alpha ^4}}-\sin ^2\left(\frac{\tau }{2 \alpha }\right)} + O(\nicefrac{1}{\sqrt{\nu}}).
\ee
The correction term is complicated, and we come to it later.
For the leading term, we see that the argument $\nicefrac{\tau}{2\alpha}$ of sine and cosine varies on a much faster time-scale than the exponential $e^{\frac{\tau ^2}{16 \alpha ^4}}$, and can be optimized independently. It is easy to see that the QFI is always maximized at $\cos ^2\left(\frac{\tau }{2 \alpha }\right)=1$. Tuning the time to ensure this,  we obtain a very simple expression
\be\label{eq: QFI alpha sq}
{\rm QFI}[\rho^{(g/e)}_{\tau|\alpha}] = \frac{\tau^2}{4 \, \alpha^4} e^{-\frac{\tau ^2}{16 \alpha ^4}} \leq e^{-1} \approx 1.47,
\ee
saturated around $\tau \approx 4 \, \alpha^2$, where the QFI oscillates fast between 0 and the maximal value.\\

Setting $\cos ^2\left(\frac{\tau }{2 \alpha }\right)=1$, the first order correction also simplifies
\begin{align}
{\rm QFI}[\rho^{(g/e)}_{\tau|\alpha}] &= \frac{\tau^2}{4\, \alpha^4} e^{-\frac{\tau ^2}{16 \alpha ^4}} - \frac{\tau }{\alpha ^2}  e^{-\frac{\tau ^2}{32 \alpha ^4}}\,  \delta \dot x.
\end{align}
The expression of $\delta \dot x$ involves the oscillating terms $\sin( \frac{\tau^2}{2\pi \nu}-\frac{\pi}{4})$ that vary quickly on a timescale of $\frac{1}{\alpha}$ and are different for all involved $\nu$. To conclude this discussion, we simply note that at finite $\alpha$ the expression of QFI in Eq.~\eqref{eq: QFI alpha sq} is modified by an oscillating contribution that only vanishes with $\nicefrac{1}{\sqrt{\alpha}}$. These oscillations are well visible in the numerical simulation for $\alpha=100$ of Fig.~\ref{fig: new}.

\subsubsection{Quantum Fisher information at very long times $\tau= O(\alpha^3)$}

For very long times $\tau \gg \alpha^2$ using Eq.~\eqref{eq: channel very long} we find (for any real intial state)
\begin{align}
    x^{(g/e)}_{\tau|\alpha}&=\frac{1}{\sqrt{3 \pi \nu}}  e^{-F^2_\nu(\tau|\alpha)} (-1)^\nu\sin(\Omega_\nu(\tau)), \qquad
    z^{(g/e)}_{\tau|\alpha}=0,
\end{align}
with $F_\nu(\tau,\alpha)=\frac{\tau^{1/3}-2 \alpha (\pi  \nu)^{1/3}}{\sqrt 2 (\pi  \nu)^{1/3}}$ and $\Omega_\nu(\tau)=\frac{3}{2} (\pi \nu \tau^2)^{1/3} +\frac{\pi}{4}$, and its derivative
\begin{equation}
    \dot x^{(g/e)}_{\tau|\alpha}= \frac{2\sqrt{2} \,F_\nu(\tau|\alpha) }{\sqrt{3 \pi \nu}}  e^{-F^2_\nu(\tau|\alpha)} (-1)^\nu\sin(\Omega_\nu(\tau)) = 2\sqrt{2} F_\nu(\tau|\alpha)  \,  x^{(g/e)}_{\tau|\alpha}.
\end{equation}
The QFI is then found to be
\be
{\rm QFI}[\rho^{(g/e)}_{\tau|\alpha}] = \frac{(\dot x^{(g/e)}_{\tau|\alpha})^2}{1- (x^{(g/e)}_{\tau|\alpha})^2}= \frac{8 F^2_\nu(\tau|\alpha)}{\nicefrac{1}{(x^{(g/e)}_{\tau|\alpha})^2}- 1} = \frac{8 F^2_\nu(\tau|\alpha)}{\frac{3 \pi \nu \, e^{2 F_\nu^{2}(\tau|\alpha)}}{\sin^2 \Omega_\nu} - 1}.
\ee
Again, the phase $\Omega_\nu$ changes on a faster scale than $F_\nu(\tau|\alpha)$ and can be optimized independently. Varying the phase, we see that the QFI oscillates between zero and the maximal value of 
\be
{\rm QFI}[\rho^{(g/e)}_{\tau|\alpha}] = \frac{8 F^2_\nu(\tau|\alpha)}{3 \pi \nu \, e^{2 F_\nu^{2}(\tau|\alpha)}- 1},
\ee
very similar to Eq.~\eqref{eq: QFI at revivals}. Close to a fixed revival, we can choose $F_\nu^{2}(\tau|\alpha)$ by varying the time. Choosing the optimal value, we find that the maximal value 
\be
{\rm QFI}[\rho^{(g/e)}_{\tau|\alpha}] = -8 \, W_0\left(-\frac{1}{3 e \pi  \nu }\right) \approx \frac{0.31}{\nu}
\ee
is attained for $F^2_\nu(\tau|\alpha)= 1+W_0\left(-\frac{1}{3 e \pi  \nu }\right) \approx 1-\nicefrac{0.04}{\nu}$, that is for $\tau \approx \pi  \nu  \left(2 \alpha \pm \left(2-\frac{0.039}{\nu }\right)\right)^3$. Such revivals of the atomic QFI at very long times are well recognizable in the numerical simulation for $\alpha=100$.

\section{Optimal successive interaction}

\label{app: optimal succ int}

In this section, we derive a successive interaction model, where the two-level probe saturates the ${\rm QFI}= 4 N$ of the coherent modes $\ket{\alpha}^{\otimes N}$.  

Before, it will be insightful to relax the requirement that the interaction is sequential. Consider the two-dimensional space spanned by the two orthonormal states
\begin{align}
\ket{\alpha_N} := \ket{\alpha}^{\otimes N} \qquad \ket{\dot \alpha_N} := \frac{(\id- \ket{\alpha_N}) \frac{\dd}{\dd \alpha }\ket{\alpha_N} }{\left\|(\id- \ket{\alpha_N}) \frac{\dd}{\dd \alpha }\ket{\alpha_N}\right\|} = \frac{1}{\sqrt{N}}\big( \ket{\dot\alpha} \ket{\alpha}^{\otimes (N-1)}+\dots + \ket{\alpha}^{\otimes (N-1)}\ket{\dot\alpha}\Big)
\end{align}
The QFI of the coherent modes is encoded in the two-dimensional subspace spanned by these states, with  $\ket{\alpha+ \dd\alpha}^{\otimes N} = \ket{\alpha_N} + \dd \alpha \sqrt{N} \ket{\dot \alpha_N} + O(\dd \alpha^2)$. It can be mapped onto the two-level probe with the help of the SWAP interaction (as in Eq.~\ref{eq: V opt})
\be
V^{(N)} = \ketbra{g,\alpha_N}{g,\alpha_N}+  \ketbra{e,\alpha_N}{g,\dot \alpha_N} +  \ketbra{g,\dot \alpha_N}{e,\alpha_N} +\ketbra{e,\dot \alpha_N}{e,\dot \alpha_N}.
\ee
For $\ket{\alpha+\dd \alpha}^{\otimes N}$, this unitary maps any initial state of the atom into $\ket{g}+ \dd \alpha \sqrt{N}\ket{e}$, whose QFI is $4 N$. However, the mapping only works well around a fixed $\alpha$,
as the coherent modes in a different state $\ket{\beta}^{\otimes N}$ quickly become orthogonal to the two-dimensional subspace described above. Moreover, this effect is amplified with $N$ as $\braket{\alpha_N}{\beta_N}=\braket{\alpha}{\beta}^N = e^{-N(\alpha-\beta)^2/2}$.\\

Based on this discussion, one can also construct a sequential interaction model achieving the same QFI. To do for the first interaction step, we set
\be
V_1 = V =\id -\ketbra{g,\dot \alpha} - \ketbra{e, \alpha}{e, \alpha}+ \ketbra{e,\alpha}{g,\dot \alpha} +  \ketbra{g,\dot \alpha}{e,\alpha} 
\ee
This maps the state $\ket{\alpha + \dd \alpha}=\ket{\alpha}+\dd \alpha\ket{\dot \alpha}$ of the field mode into $\ket{g}+ \dd \alpha \ket{e}$. Right before the second interaction, the state of the atom and the second mode is thus
\be
\Big(\ket{g}+\dd \alpha \ket{e}\Big) \otimes \Big(\ket{\alpha}+\dd \alpha \ket{\dot \alpha}\Big) = \ket{g,\alpha} + \dd \alpha(\ket{e, \alpha}+\ket{g,\dot \alpha}) + O(\dd \alpha^2).
\ee
It is now the two-dimensional subspace $\left\{ \ket{g,\alpha} ,\ket{\psi_2} :=\frac{1}{\sqrt 2}(\ket{e, \alpha}+\ket{g,\dot \alpha})\right \}$ that carries the QFI of the global state. This subspace can again be swapped into the probe by means of the unitary interaction
\be
V_2 = \id -\ketbra{g,\dot \alpha}- \ketbra{e,\alpha}+  \ketbra{e,\alpha}{ \psi_2} + \ketbra{g,\dot\alpha}{ \psi_2^\perp}
\ee
with $\ket{\psi_2^\perp} :=\frac{1}{\sqrt 2}( \ket{g,\dot \alpha}-\ket{e, \alpha})$, preparing the atomic state in $\ket{g}+ \dd \alpha \sqrt{2} \ket{e}$. \\

This construction can be generalized to all steps. Using the appropriate interaction 
\be
V_k =\id -\ketbra{g,\dot \alpha}- \ketbra{e,\alpha}+  \ketbra{e,\alpha}{ \psi_k} + \ketbra{g,\dot\alpha}{ \psi_k^\perp}
\ee
with $\ket{\psi_k}= \frac{1}{\sqrt k+1}(\sqrt{k} \ket{e,\alpha}+ \ket{g,\dot\alpha})$ and $\ket{\psi_k^\perp}= \frac{1}{\sqrt k+1}(\ket{e,\alpha}-\sqrt{k}  \ket{g,\dot\alpha})$,  one can always swap the appropriate global qubit subspace into the probe
\be
V_k: \ket{g,\alpha} +  \dd \alpha (\sqrt{k} \ket{e,\alpha}+ \ket{g,\dot\alpha} ) \quad\mapsto \quad \ket{g,\alpha} +\dd \alpha \sqrt{k+1} \ket{e,\alpha},
\ee
ensuring that the probe's QFI is given by $4k$ at all steps.
Hence, applying the successive interactions $V_1,\dots, V_N$ with a series of coherent modes $\ket{\alpha}$, leads to the marginal state of the probe with ${\rm QFI}=4N$ saturating the upper bound. Importantly, the interactions $V_k$ are step-dependent and fine-tuned to work around a specific value of $\alpha$.

\section{The infinite modes limit}

\label{app: infinite modes}
In this section, we briefly discuss the limit $\tau=\dd t\to 0$ and $\alpha=\it cst$. Here, we obtain for the channel matrix
 \begin{align}
G = \left(\begin{array}{cccc}
 1-\tau^2 \alpha^2 & 1- \tau^2(\alpha^2 +\frac{1}{2}) & -\tau \alpha & \tau\alpha \\
1- \tau^2(\alpha^2 +\frac{1}{2}) & 1-\tau^2(\alpha^2 +1) & -\tau \alpha & \tau \alpha\\
 -\tau \alpha & -\tau \alpha   & \tau^2 (\alpha^2+1) &-\tau^2 \alpha^2\\
 \tau \alpha & \tau\alpha &-\tau^2 \alpha^2 & \tau^2 \alpha^2
\end{array}\right)
=
\left(\begin{array}{cccc}
 1 & 1 & -\alpha \dd t & \alpha \dd t \\
1 & 1 & -\alpha \dd t & \alpha \dd t \\
 -\alpha \dd t  & -\alpha \dd t    & 0  &0\\
 \alpha \dd t  & \alpha \dd t  & 0& 0
\end{array}\right) + O(\dd t^2)
\end{align}
so the atomic dynamics is unitary
\be
 \frac{\dd }{\dd t} \rho_{t|\varepsilon} = - \ii \, \alpha\,  [\sigma_y,\rho_{t|\varepsilon} ].
 \ee

\section{The continuous quantum field limit}
\label{app:collision}

In this section, we compute the limit where the interaction parameter for a single discrete mode is $\tau = \sqrt{  \kappa \dd t}$ and the number of photons per mode is $\bar n = \varepsilon^2 \dd t$, where $\varepsilon^2$ is the photon flux and the dimensionless field amplitude $\varepsilon$ is our estimated parameter.

First, let us show that in this limit one recovers the atom coupled to a bosonic field models, studied in e.g. \cite{dkabrowska2021eternally,Fischer2018scatteringintoone,Schafer_2021}. For a single mode, the propagator reads 
\begin{equation}
    U_\tau^{(t)} = \exp(\tau (a_t \sigma_+ -a_t^\dag \sigma_-)) = \exp(\sqrt{\kappa} (\sqrt{\dd t} a_t \sigma_+ -\sqrt{\dd t} a_t^\dag \sigma_-))
\end{equation}
Taking the limit $\dd t \to 0$, one notices that the rescaled operators $a(t)= \frac{a_t}{\sqrt{\dd t}}$ satisfy the commutation relations of a bosonic field $[ a(t),a^\dag(t')] = \frac{\delta_{t,t' }}{\dd t}\to \delta(t-t').$ Hence, we have
\begin{equation}
    U_\tau^{(t)} = \exp(\sqrt{\kappa} ( a(t) \sigma_+ - a^\dag(t) \sigma_-)\dd t) =  1 -\ii\, \dd t \, H(t) + O(\dd t^2)
\end{equation}
with the time-dependent Hamiltonian $H(t) = \ii \sqrt{\kappa} ( a^\dag(t) \sigma_- -  a(t) \sigma_+ )$ coupling the atom to the the quantum field. Next, consider the initial state of the field modes. By assumption, the discrete time modes $a_t$ are initially prepared in the coherent states $\ket{\alpha}= \mathcal{D}(\alpha)\ket{0}$ with $\alpha^2 = \bar n = \varepsilon^2 \dd t$. The unitary disparagement operators $\mathcal{D}(\alpha)$ transform the creation annihilation operators as 
\be
\mathcal{D}^\dag(\alpha) a_t \mathcal{D}(\alpha) = a_t +\alpha \implies \mathcal{D}^\dag(\alpha) a(t) \mathcal{D}(\alpha) = a(t) +\frac{\alpha}{\sqrt{\dd t}} = a(t) + \varepsilon.
\ee
We can rewrite the interaction Hamiltonian with the expression given in the main text as
\be
H(t) = \ii\,  \sqrt{\kappa} \Big(\sigma_- (\varepsilon + a^\dag(t)) - \sigma_+ (\varepsilon+a(t))\Big),
\ee
with $a(t),a^\dag(t)$ now describing a vacuum quantum field.

This expression is, however, not needed to derive the reduced dynamics of the atom. Rather, we take the expression \eqref{eq: channel in Gramm} of the CPTP map 
$\cE_{\tau|\mathfrak{F}}[\bullet]$ and expand it for a short times $\tau=\sqrt{\kappa \dd t} \to 0$. Assuming that the state has bounded energy $\tau \sqrt{\hat n} \ll 1$, we expand the trigonometric function in the leading and  obtain for the Gram matrix
\be
G \approx \tilde G = 
\left(\begin{array}{cccc}
 1-\tau^2 \mean{\hat n}  & 1- \tau^2(\mean{\hat n} +\frac{1}{2}) & -\tau \mean{a^\dag} & \tau\mean{a} \\
1- \tau^2(\mean{\hat n} +\frac{1}{2}) & 1-\tau^2(\mean{\hat n} +1) & -\tau \mean{a^\dag} & \tau\mean{a} \\
 -\tau \mean{a} & -\tau\mean{a}    & \tau^2 (\mean{\hat n}+1) &-\tau^2 \mean{a^{\dag 2}}\\
 \tau \mean{a^\dag} & \tau\mean{a^\dag} &-\tau^2 \mean{a^2} & \tau^2 \mean{\hat n}
\end{array}\right),
\ee
where the expected values $\mean{A} = \tr A \ketbra{\mathfrak{F}}$ are taken on the initial state of the incident radiation mode. Taking the field mode in the coherent state $\ket{\alpha}=\ket{\varepsilon\sqrt{\dd t}}$, we further obtain $G = \tilde G + \mathcal{O}((\tau \alpha)^3)$ with
\begin{align}
\tilde G = \left(\begin{array}{cccc}
 1-\tau^2 \alpha^2 & 1- \tau^2(\alpha^2 +\frac{1}{2}) & -\tau \alpha & \tau\alpha \\
1- \tau^2(\alpha^2 +\frac{1}{2}) & 1-\tau^2(\alpha^2 +1) & -\tau \alpha & \tau \alpha\\
 -\tau \alpha & -\tau \alpha   & \tau^2 (\alpha^2+1) &-\tau^2 \alpha^2\\
 \tau \alpha & \tau\alpha &-\tau^2 \alpha^2 & \tau^2 \alpha^2
\end{array}\right)
=
\left(\begin{array}{cccc}
 1 & 1- \frac{\kappa \dd t}{2} & -\sqrt{\kappa} \varepsilon \dd t & \sqrt{\kappa} \varepsilon \dd t \\
1- \frac{\kappa \dd t}{2} & 1- \kappa \dd t & -\sqrt{\kappa} \varepsilon \dd t & \sqrt{\kappa} \varepsilon \dd t \\
 -\sqrt{\kappa} \varepsilon \dd t  & -\sqrt{\kappa} \varepsilon \dd t    & \kappa  \dd t  &0\\
 \sqrt{\kappa} \varepsilon \dd t  & \sqrt{\kappa} \varepsilon \dd t  & 0& 0
\end{array}\right),
\end{align}
where we only kept the leading orders in $\dd t$ in the last expression. It is then straightforward to verify that the resulting channel is of the form 
\be
\cE_{\dd t|\varepsilon}[\bullet] = \sum_{i,j=0}^3 G_{ji} \, L_i \bullet L_j^\dag, =\bullet + \dd t\left(- \ii \sqrt{\kappa} \,\varepsilon [\sigma_y,\bullet] + \kappa \, \mathcal{L}_{\sigma_-}[\bullet]\right) + O(\dd t^2), 
\ee
  with the dissipator $\mathcal{L}_{A}[\bullet] = A \bullet A^\dag- \frac{1}{2} \{A^\dag A , \bullet \}$ (and  $\{L_0,L_1,L_2,L_3\} =\{ \ketbra{g},\ketbra{e},\sigma_-,\sigma_+\}$). The reduced dynamics of the atom are thus governed by the master equation
  \be
 \frac{\dd }{\dd t} \rho_{t|\varepsilon} = - \ii \sqrt{\kappa} \,\varepsilon \,  [\sigma_y,\rho_{t|\varepsilon} ] + \kappa \, \mathcal{L}_{\sigma_-}[\rho_{t|\varepsilon}],
 \ee
  featuring coherent driving and spontaneous emission.\\


\section{QFI for $\bar{\varepsilon}=0$}\label{app: QFI-epsilon-bar}
In this section, we give the explicit formula of the QFI for any initial state $\rho_0=\frac{1}{2}(\id+x_0\sigma_x+z_0 \sigma_z)$ which dynamics are given in Eq.~\eqref{eq: master} for fixed $\bar{\varepsilon}=0$. It reads

\be
{\rm QFI}[\rho_{s|\bar \varepsilon=0}]=-\frac{16 e^{-s} \left(e^{s/4}-1\right)^2 \left(-\left(e^{s/4} \left(x_0^2+z_0-1\right)+e^{s/2} (z_0-1)+e^{\frac{3 s}{4}}+(z_0-1)^2\right)^2+e^{s} x_0^2+e^{s} \left(e^{s/4}+z_0-1\right)^2\right)}{e^{s/2} \left(x_0^2+2 z_0-2\right)+(z_0-1)^2}.
\ee

\end{document}